%% file: main.tex
\documentclass[twocolumn,citeautoscript,pra,superscriptaddress,amsmath,amssymb]{revtex4-1}

\usepackage{graphicx}
\usepackage{color}
\usepackage{upgreek}
\usepackage{float}
\usepackage{lipsum}
\usepackage{graphicx}
\usepackage{color}
\usepackage{upgreek}
\usepackage{float}
\usepackage{booktabs}
\usepackage{amsfonts}
\usepackage{amsmath}
\usepackage{amssymb}
\usepackage{makecell}
\usepackage{mathrsfs}
\usepackage{fancyhdr}
\usepackage{mathtools}
\usepackage{subcaption}
\usepackage{enumitem}
\usepackage[table,xcdraw]{xcolor}
\usepackage[utf8]{inputenc}
\usepackage[english]{babel}
\usepackage[breaklinks,colorlinks,citecolor=blue,linkcolor=blue]{hyperref}
 
\newtheorem{theorem}{Theorem}
\newtheorem{corollary}{Corollary}[theorem]

\newcommand{\ket}[1]{\ensuremath{\left| #1 \right\rangle}}

\AtBeginDocument{
\heavyrulewidth=.08em
\lightrulewidth=.05em
\cmidrulewidth=.03em
\belowrulesep=.65ex
\belowbottomsep=0pt
\aboverulesep=.4ex
\abovetopsep=0pt
\cmidrulesep=\doublerulesep
\cmidrulekern=.5em
\defaultaddspace=.5em
}

\begin{document}

\title{Truncated phase-based quantum arithmetic: error propagation and resource reduction}

\author{G. A. L. White}
\email{greg.al.white@gmail.com}
\affiliation{School of Physics, University of Melbourne, Parkville, VIC 3010, Australia}

\author{C. D. Hill}
\affiliation{School of Physics, University of Melbourne, Parkville, VIC 3010, Australia}
\affiliation{School of Mathematics and Statistics, University of Melbourne, Parkville, VIC, 3010, Australia}

\author{L. C. L. Hollenberg}
\affiliation{School of Physics, University of Melbourne, Parkville, VIC 3010, Australia}

\begin{abstract}
There are two important, and potentially interconnecting, avenues to the realisation of large-scale quantum algorithms: improvement of the hardware, and reduction of resource requirements demanded by algorithm components. 
In focusing on the latter, one crucial subroutine to many sought-after applications is the quantum adder.
A variety of different implementations exist with idiosyncratic pros and cons. 
One of these, the Draper quantum Fourier adder, offers the lowest qubit count of any adder, but requires a substantial number of gates as well as extremely fine rotations. In this work, we present a modification of the Draper adder which eliminates small-angle rotations to highly coarse levels, matched with some strategic corrections. This reduces hardware requirements without sacrificing the qubit saving. 
We show that the inherited loss of fidelity is directly given by the rate of carry and borrow bits in the computation. We derive formulae to predict this, complemented by complete gate-level matrix product state simulations of the circuit. Moreover, we analytically describe the effects of possible stochastic control error.
We present an in-depth analysis of this approach in the context of Shor's algorithm, focusing on the factoring of RSA-2048. Surprisingly, we find that each of the $7\times 10^7$ quantum Fourier transforms may be truncated down to $\pi/64$, with additive rotations left only slightly finer. This result is much more efficient than previously realised. We quantify savings both in terms of logical resources and raw magic states, demonstrating that phase adders can be competitive with Toffoli-based constructions.
\end{abstract}

\maketitle
\input{intro.tex}

\input{analytic_trunc.tex}

\input{trunc_depth.tex}

\input{error_model.tex}

\input{adder_redesign.tex}

\input{cost_estimates.tex}

\input{conclusion.tex}

%

\clearpage
\onecolumngrid
\appendix
\input{appendices.tex}

\input{LNN_Shor.tex}

\end{document}

%% file: intro.tex
\section{Introduction}
\label{sec:intro}

While the field of theoretical quantum computing is still relatively new, progress in this area has produced some highly enticing results, with even a modest number of envisaged applications spurring the race to build a universal quantum computer~\cite{Nielsen2010QuantumInformationb,RevModPhys.82.1,montanaro2016quantum}. 
However, discussions of quantum algorithms typically take place in the abstract: the low-level underpinnings are hidden in a black-box framework.
This leaves a great deal of room for circuit optimisation within each component -- in particular, which operations are necessary at a practical level.
Indeed, for quantum computers to realise their full potential, it is crucial that circuit design meet hardware advances in the middle. Quantum resources such as circuit size, depth, gate count, and fault tolerant resources in particular must be kept at a minimum. \par
A circuit component key to many applications of a quantum computer is the adder.
Basic arithmetic operations are anticipated to be crucial to many useful quantum algorithms, as in the classical case~\cite{Li2020,Ruiz-Perez2017,babbush2016exponentially,harrow2009quantum,asaka2020quantum}. The most famous use-case is that of the modular exponentiation performed in Shor's algorithm.
There are two fundamental methods in order to achieve this.
The first method uses Toffoli gates that mirror classical binary compositions \cite{gidney-addition, cuccaro-adder, eth-adder,gidney2021factor}.
The second method for performing arithmetic is an inherently quantum routine.
Known as the Draper adder \cite{draper-adder, Pavlidis2014,Ruiz-Perez2017}, this involves the application of a quantum Fourier transform (QFT), a sequence of structured $Z-$rotations to each qubit, followed by an inverse QFT (IQFT).
Known constants can be semi-classically added to a quantum register this way through rotations, which, after transformation back to the computational basis, correspond to a displacement by a fixed number.
Since the numbers do not need to be stored in a second register, this method halves the number of required qubits, making it especially appealing for near-term applications. 
The basic approach has drawbacks both in resource requirements and demands of extraordinarily fine phase precision.  

In this work, we make several contributions to the study of phase-based arithmetic. We examine the effects of eliminating gate rotations below some fixed level $\pi/2^\mathcal{N}$ within a quantum adder, showing that the error induced depends entirely on the numbers being summed. We derive both the resulting exact and the average-case loss in fidelity, finding a remarkable robustness to truncation. Using our analysis, we modify the quantum adder to include some informed corrective rotations at no additional gate cost, permitting far coarser truncations even for extremely large or repeated components. Reducing the requirements of arithmetic brings large-scale quantum algorithms a step closer.
Adapting these tools, we also investigate analytically and numerically the effects of basic stochastic control errors in the phase rotations. We provide rigorous estimates for required tolerances in these gates, supplemented by numerical simulations. This analysis is essential for implementation not only in the NISQ era, but also in the far-term where non-Clifford gates are expected to dominate error rates. \par
Our investigation considers both standalone arithmetic components, and in the context of a Shor's algorithm circuit -- with the particular focus on the factoring of RSA-2048. In particular, for Shor's algorithm targeting RSA-2048, we study the circuit of Ref.~\cite{Fowler2004b} and find that QFT rotations can be removed up to $\pi/64$ -- a three order of magnitude reduction in gates. This is surprising, because for an $L$ bit number there are $\mathcal{O}(16L^2) \approx 7\times 10^7$ QFTs in Shor's algorithm. Ostensibly one might expect that removal of rotations up to $\pi/128 \approx 0.025$ would produce an error of this magnitude in each of the $\mathcal{O}(8L^4)$ locations, catastrophic to the computation. We show that the interplay of different components is far more structured, preventing errors from necessarily multiplying out and compromising the algorithm.

Our approach eliminates the majority of the logical resources required of the circuit. Although truncation does not reduce circuit depth, it reduces the effects of gate error. Further, limiting the fineness of rotations reduces the cost of gate synthesis in fault tolerant contexts.
We show that for a standard implementation of Shor's algorithm for RSA-2048, this method consumes around an order of magnitude more raw magic states than the most optimal Toffoli-based construction, but with 2044 fewer logical qubits.\par 

Previous work in the literature has covered the idea of an approximate QFT (AQFT) through the removal of finer rotation gates by studying the effects on the operator itself \cite{coppersmith-est,Nielsen2010QuantumInformationb,PhysRevA.54.139,woolfe-mps,PhysRevA.76.042321} -- concluding usually that only the exponentially small components should be omitted (comparable to, for example, the tolerable noise level). Moreover, Refs.~\cite{Fowler2004,PhysRevA.87.032333} explore the coarser truncation of the QFT in the context of period-finding in Shor's algorithm.
We provide an account in arithmetic of how the removal of such rotation gates explicitly depends on the numbers being added. Since the underlying probability distribution is typically known, for example in the initialisation of a register in equal superposition, the average-case performance can be determined. Due to factors such as infrequent error occurrence and natural error cancellation, we find, surprisingly, that the phase adder is far more robust to deliberate truncation than previously established. \par 
In short, we relate the exact error incurred through truncation to the frequency and significance of carry and borrow bits in the addition and subtraction of binary numbers. Using this, we show how interleaving addition and two's complement subtraction (negative rotation) circuits can cancel most errors, and provide a surprising natural robustness to errors in arithmetic components despite the removal of most gates. The results we arrive at are simple to compute, and match numerical simulations precisely.\par

The paper is organised as follows: In Section~\ref{sec:background} we explicitly comb through the structure of the Draper adder, the understanding of which is central to the remainder of the paper.
In Section~\ref{sec:analytic-trunc}, we derive an expression for the exact effects of truncation in quantum arithmetic by removing all rotation gates finer than some given angle.
Using this, we then compute the consequential error incurred on an average quantum circuit for both small and asymptotically large $L$.
Next, we generalise the investigation to multiple instances of an adder, combining into higher levels of arithmetic in Section~\ref{sec:trunc-depth}. 
To address the practicalities of error-prone quantum computing, we construct a circuit error model combining $Z-$rotation errors with surface code language, and derive an analytic model for the performance of arithmetic components under these generic noise models in Section~\ref{sec:error-model}.
Using these key pieces of information, in Section~\ref{sec:adder-redesign} we then propose a redesign of the quantum adder which eliminates unnecessary phase gates without sacrificing the computation.\par 
We examine a case study of this truncated adder in the context of Shor's algorithm, and its performance there -- the results of which can be found in Sections~\ref{sec:adder-redesign} and~\ref{sec:cost-analysis}, but the details of which can be found in Appendices~\ref{sec:LNN-shor},~\ref{app:shor-trunc}, and~\ref{app:shor-error}. Our results are shown in context of the circuit from Ref.~\cite{Fowler2004b}, but could be straightforwardly adapted to other resource efficient approaches, such as in Ref.~\cite{Pavlidis2014}.
Finally, in Section~\ref{sec:cost-analysis}, we analyse the circuit costs involved in the context of the surface code, and make comparison to other addition circuits in the literature. 
In particular, this is in terms of raw $T$ states consumed in magic state distillation~\cite{magic-distillation-campbell}.\par
To supplement the arguments made in this work we used a matrix product state (MPS) simulator from Ref.~\cite{dang-mps} at the gate level to obtain the exact quantum states in the relevant circuits. For truncated arithmetic components, this is up to 60 qubits. 
We derive expressions for the exact effects of truncation; compare these with the MPS results; evaluate the average effects; compare these with the average MPS results; and finally compare the average results to a Monte Carlo simulation of the correct state probability in the limit of large $L$. 
Some mathematical approximations are made in the derivation of our scaling formulae, so the agreement with simulation is crucial to these arguments.\par 
\begin{figure*}
\centering
\includegraphics[width=\linewidth]{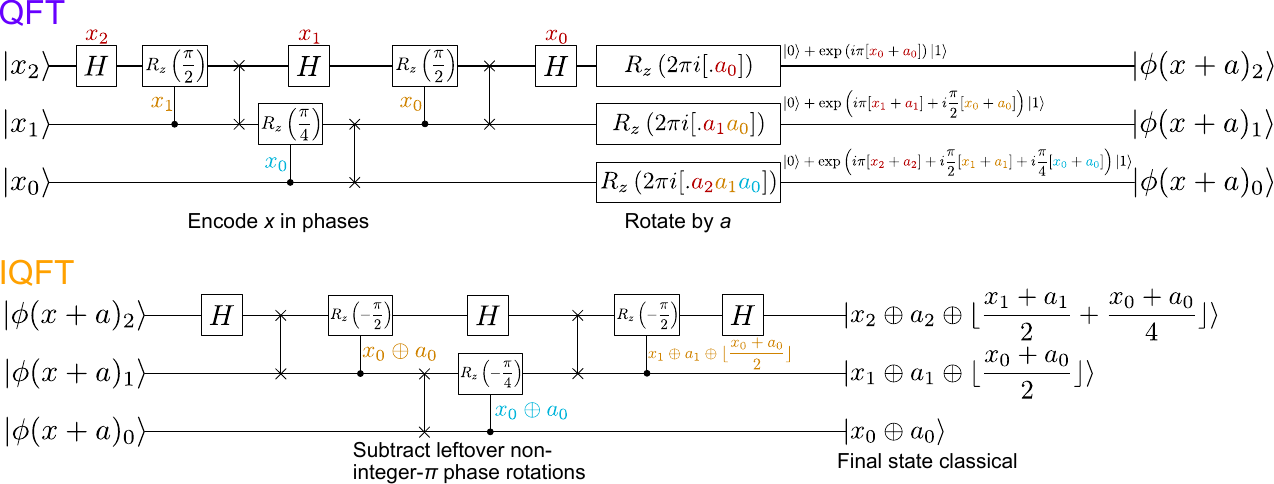}

\caption{Circuit diagram of a three-qubit Draper adder. A basis state $x$ is encoded into the phases of each qubit using the QFT. Some number $a$ is then added (modulo 8) by performing $Z$-rotations for each $a_i$ at the equivalent angle by which $x_i$ is encoded (here: depicted with the same colour). The controlled rotations in the IQFT then subtract off any non-carry bits, and finally a Hadamard transforms all states of the form $\ket{0} + \exp(i\pi\cdot y)\ket{1}$ into the definite state $\ket{y}$. Importantly, the necessity of the rotation gates depends on the numbers that are added.}
\label{draper}
\end{figure*}
\section{An overview of the Draper adder and its truncation}
\label{sec:background}

Understanding of the fine-workings of the Draper adder is key to arguments made in this work.
In this section, we will outline the basic operation of phase-based quantum arithmetic, as well as the philosophy behind truncation and previous work.\par 
The mechanics of addition of a number $a$ on a quantum register follow the steps laid out in Figure \ref{draper}.
That is, after a QFT transforms a register into the Fourier basis, a sequence of additive rotations can be performed that corresponding to addition in the computational basis.\par

It is important for the remainder of this manuscript to scrutinise exactly how this adder functions at the bit-level.
For the addition of two $L$-bit numbers $x$ and $a$, a QFT is performed on the register $\ket{x}$. 
In the factorised form each $i$th qubit has a relative phase of $\text{e}^{2\pi i(.x_i\dots x_0)}$, where we use the little-endian convention. The least significant bit is denoted $x_0$, and stored in the bottom-most qubit register.
Following this are the necessary $a$ rotations on each qubit, delivering $\ket{\phi(x+a)}$ in the factorised form:

\begin{equation}
\label{factor-form}
\bigotimes_{j=0}^L\left(\ket{0} + \exp\left[\sum_{i=0}^j 2\pi i\left(\frac{x_i}{2^i} + \frac{a_i}{2^i}\right)\right]\ket{1}\right).
\end{equation}

This will be transformed back into the computational basis with the application of an IQFT.
This acts sequentially from left to right on each of these qubits.
On $$\ket{\phi(x+a)_{n-1}} = (\ket{0} + \text{e}^{2\pi i(.x_0 + .a_0)}\ket{1})$$ a Hadamard will operate.
If $x_0 + a_0 = 1$, then $$\text{e}^{2\pi i(.x_0 + .a_0)}=\text{e}^{i\pi}=-1$$ and $H(\ket{0}-\ket{1})/\sqrt{2} = \ket{1}$.
Similarly, if $x_0+a_0=0$, then a Hadamard will deliver the state $\ket{0}$.
However, if $x_0+a_0 = 2$, then $$\text{e}^{2\pi i(.x_0 + .a_0)}=\text{e}^{2\pi i}=1$$ and $H(\ket{0}+\ket{1})/\sqrt{2} = \ket{0}$.
For subsequent usage, we denote this qubit state $\ket{y_0}$.
The next qubit, $$\ket{\phi(x+a)_{n-2}}=(\ket{0} + \text{e}^{2\pi i(.x_1x_0 + .a_1a_0)}\ket{1}).$$
Firstly, we have a $-\pi/2$ rotation which is controlled by $\ket{y_0}$.
This transforms the state into $$(\ket{0} + \text{e}^{2\pi i(.x_1x_0 + .a_1a_0 - .0y_0)}\ket{1}).$$
If $x_0 = a_0 = 0$, the state will be $$(\ket{0} + \text{e}^{2\pi i(.x_1+ .a_1)}\ket{1}).$$
Similarly, if $x_0+a_0 = 1$, then $x_0 + a_0 - y_0 = 1 - 1 = 0$ and the state will still be $$(\ket{0} + \text{e}^{2\pi i(.x_1 + .a_1)}\ket{1}).$$
However, if $x_0 + a_0 = 2$ (i.e. $x_0 \oplus a_0 = 0$), $\ket{y_0}$ will \textit{not} control this rotation.
The state is then left as $$(\ket{0} + \text{e}^{2\pi i(.x_11 + .a_11)}\ket{1}) = (\ket{0} + \text{e}^{2\pi i(.x_1 + .a_1 + .1)}\ket{1})$$ -- the carry bit travels to the next level of significance for free.
This is the elegance of the Draper adder.
All numerical information is progressively built up on each qubit, and IQFT subtracts away extra phase rotations to ensure that it ends as an integer multiple of $\pi$.
A Hadamard will then transform $\ket{\phi(x+a)_{n-2}}$ into $\ket{1}$ if $x_1=a_1=0$; $\ket{0}$ if $x_1 + a_1 = 1$, and $\ket{1}$ if $x_1 + a_1 = 2$.
In the latter two cases, the carry bit will propagate further in the same fashion.
Once all of the inverse phase rotations are completed, all carry bits are in the correct position.
The $j^\text{th}$ qubit will be found in the state $$H(\ket{0}+\text{e}^{i\pi y_j}\ket{1})/\sqrt{2} = \ket{y_j}$$ \par
Generally, each qubit is in the form
\begin{equation}
\frac{1}{2}\left[\left(\text{e}^{i\theta}+1\right)\ket{0} + \left(\text{e}^{i\theta}-1\right)\ket{1}\right],
\end{equation}
where $\theta = 2\pi (.x_j\cdots x_0 + .a_j\cdots a_0 -.0y_{j-1}\cdots y_0)$ for a perfectly functioning adder, however as we will see, the angle may also depend on how truncation is performed.
The probability of having a final register $\ket{y_{L-1}\cdots y_0}$ is therefore the absolute value squared of the product of each of these qubit amplitudes:
\begin{widetext}
\begin{equation}
\label{adder-expression}
    \begin{split}
        \text{Pr}\left(y_{L-1}\cdots y_0\right)&= |\langle y_{L-1}\cdots y_0|\psi\rangle|^2\\
        & = \left| \frac{1}{2^L} \prod_{j=0}^{L-1}\left[\left(-1\right)^{y_j} + \exp\left(i\sum_{k=0}^j 2\pi \left(\frac{x_k + a_k}{2^{j-k+1}}\right)-\sum_{k=1}^{j}\left(\frac{i\pi y_{k-1}}{2^{j-k+1}}\right)\right)\right]\right|^2
    \end{split}
\end{equation}
\end{widetext}
For a perfectly functioning adder, this will be unity for the correct $y_{L-1}\cdots y_0$, and zero for all others.
We will return to this expression later in order to investigate more clearly the effects of any modifications.
The key point of this exposition is that the finer rotations are required in order to carry information from further down the register.
That is, the principal $\pi$ rotation is enough for addition of two bits at a single location, and the finer rotations deposit carry bits from earlier in the register.

Once addition is established, all other arithmetic operations can be implemented as an extension of this procedure. In particular:
\begin{itemize}
    \item \textbf{Subtraction} can be equivalently performed with negative rotations;
    \item \textbf{Multiplication} is achieved through repeated addition with the aid of an $n$-bit ancilla register initialised to zero. Each bit of a register $\ket{x_{n-1}\cdots x_0}$ controls additions of $2^j\cdot a$, adding to $x_0 2^0\cdot a + x_1 2^1\cdot a +\cdots + x_{n-1}2^{n-1}\cdot a = \left(x_{n-1}2^{n-1} + \cdots + x_12^1 + 2^0x_0\right)\cdot a = x\cdot a$;

    \item \textbf{Exponentiation} is achieved through repeated multiplication. Starting once more with an ancilla register, this time it is initialised to the state $\ket{00\cdots 01}$. In order to perform $a^x$ for some number $a$ and some quantum register $\ket{x}=\ket{x_{n-1}\cdots x_0}$ each qubit of the $\ket{x}$ register must control a multiplication of $a^{2^i}$ such that the ancilla register transforms as: $\ket{1}\mapsto \ket{1\cdot a^{2^{n-1}x_{n-1}}\cdot a^{2^{n-2}x_{n-2}}\cdots a^{2^0x_0}} = \ket{a^{2^{n-1}x_{n-1} + 2^{n-2}x_{n-2} + \cdots + 2^0x_0}} = \ket{a^x}$.
\end{itemize}

In many physical quantum computing architectures, connectivity is a limited resource. For this reason, we keep our discussions to the most restrictive case: that of linear nearest-neighbour (LNN) interactions. Moreover, non-LNN physical architectures do not preclude an LNN restriction at the logical level. Relaxations to more connective architectures requires a simple reduction of swap gates in the circuit, all other results will still be consistent. 
Our circuit model discussion of Fourier arithmetic follows the LNN circuits outlined in \cite{Fowler2004b}.
The main difference in comparison to typical QFT circuits is that the SWAP gates are interleaved between the controlled rotation gates, rather than all at the end.
The circuit diagram of the LNN QFT is shown in Figure \ref{draper}. \par
The concept of an AQFT is known in the literature \cite{coppersmith-est,Fowler2004}.
That is, the idea that rotations in the QFT become exponentially fine with register size, and can be neglected with minimal error at some point.
However, it has not been comprehensively studied in the context of Fourier arithmetic -- only in small-scale numerics \cite{PhysRevA.88.062310}.
Here, we analytically and numerically study the effects of truncating the phase rotations in the Draper adder for both small and large numbers, and use our results to redesign the structure to be significantly more resource-efficient.


%% file: analytic_trunc.tex
\section{Analytic treatment of truncation in a single addition circuit}
\label{sec:analytic-trunc}
In this section, we consider the application of a Draper adder in a single addition circuit where the QFT, rotations, and IQFT rotation gates are truncated down to a level we will denote by $\mathcal{N}$, where no rotation is more fine than $\pi/2^\mathcal{N}$.
Although truncated QFT in arithmetic is similar to the approximate QFT, a key distinction is that the error incurred depends on the input to the operation, rather than being inherent to the operation itself.
That is, the level of error will depend \emph{only} on the two numbers being added.
Equation (\ref{factor-form}) shows how, in Fourier space, the state is stored progressively on each qubit.
Each bit is attached to a $\pi/2$ phase rotation on some particular qubit.
In truncating our phase precision, there is no material loss of information about the addition.
The information is present, but no longer distributed to every qubit.
Whether this induces an error or not depends on its effect on carry bits.
For example, if the numbers 5 and 2 are added together -- or $101_2 + 010_2$ -- there will be no carry bits, and truncation as coarse as $\pi$ will suffice for the addition without error.
Using Equation (\ref{adder-expression}), we can derive an exact expression for the error incurred when the two numbers being added are known.
In the general case of a quantum algorithm, however, the register number $x$ will typically be in an equal superposition, and only the added number $a$ will be known. In general, we consider a random $a$, but with knowledge of this number the model can be updated, as will be seen later.
In this more general scenario, we derive an expression for the average error that occurs in the addition of random numbers.\par
\begin{figure*}
\centering
\includegraphics[width=0.9\linewidth]{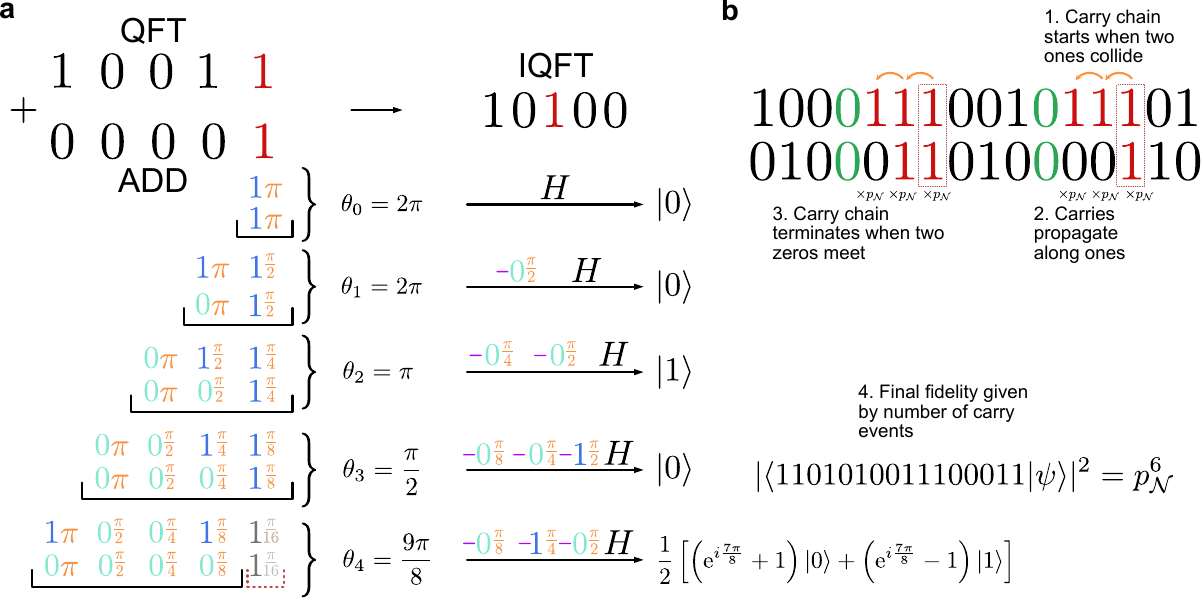}
\caption{Example error propagation in a truncated Draper circuit. \textbf{a} Adding together two five-bit numbers truncated down to $\pi/8$. A carry occurs at the least significant position and is accounted for by the first four bits. However, truncation of the $\pi/16$ phase angles means the most significant bit is not aware of the origin of the carry. Consequently, there is a mismatch where the IQFT subtracts off the resulting carry bit, but the original phases are missing, causing an overrotation by $\pi/8$. \textbf{b} At a higher level, the circuit fidelity will be multiplied by a factor of $p_\mathcal{N}$ for each carry event. Carry chains start when $x_i$ and $a_i$ are both 1, for some $i$. The number of carry bits then propagate when either or both of the subsequent $x$ and $a$ bits are 1. The carry chain terminates when $x_j$ and $a_j$ are both zero, for some $j>i$. The number of carry events is then the number of carry chains multiplied by the length of each carry chain.} 
\label{fig:trunc-main}
\end{figure*}
This is distinguished from the usual ideas concerning AQFTs, wherein the approximation is hinged on the operator rather than its input.\par 
Consider a four qubit register with $x=3 $ $(\ket{0011})$ and $a = 4 (\ket{0100})$, truncating to $\mathcal{N}=2$.
The factorised form after additive rotations is
\begin{equation}
\begin{split}
        &\left(\ket{0} + \text{e}^{2\pi i(.1+.1)}\ket{1}\right)\otimes \left(\ket{0} + \text{e}^{2\pi i(.11+.11)}\ket{1}\right) \otimes \\
        &\left(\ket{0} + \text{e}^{2\pi i(.011 + .011)}\ket{1}\right) \otimes \left(\ket{0} + \text{e}^{2\pi i(.001+.001)}\ket{1}\right).
\end{split}
\end{equation}

The first three qubits include all relevant rotations.
As a result, these remain in the respective definite states $\ket{0},\ket{1},\ket{1}$.
The last qubit, however, is in the state  $$(\text{e}^{2\pi i(.001+.001)}+1)\ket{0} +(\text{e}^{2\pi i(.001+.001)}-1)\ket{1}),$$ with controlled rotations $-\pi/4$ from the second qubit, $-\pi/2$ from the third qubit.
Thus, it is $$((\text{e}^{2\pi i(.001+.001 -.011)}+1)\ket{0} +(\text{e}^{2\pi i(.001+.001-.011)}-1)\ket{1}).$$
The truncation of the first bit means the negative phase has over-rotated, leaving this qubit as $$((\text{e}^{(-i\pi/4)}+1)\ket{0} +(\text{e}^{(-i\pi/4)}-1)\ket{1}).$$
The final state of the register is now (amplitude, disregarding phase) $\approx \sqrt{0.854}\ket{0110} + \sqrt{0.146}\ket{1110}$.
That is, truncation of size $\mathcal{N}$ with a carry bit reduces the probability of obtaining the correct result.
We will denote the remaining probability as 
\begin{equation}
    p_{\mathcal{N}} := \left|\frac{1}{2}\exp{\left(-\frac{i\pi }{2^\mathcal{N}}\right)}+\frac{1}{2}\right|^2,
\end{equation}
and refer to it as the \emph{carry fidelity}.\par
To summarise: the origin of the carry bit is omitted because of the truncation; the carry bit itself is present in the IQFT, resulting in a mismatch.
Consequently, the state is over-rotated past the origin. Figure \ref{fig:trunc-main}a steps schematically through an example of a truncation effect.\par
It is important to stress that the effects of truncation do not yield an \emph{error} as such (although we will be liberal with the term), but rather it is an intentional miscalculation.
We are investigating the exact extent to which we can modify our calculation methods and still end up with a probabilistically correct answer.
If this calculation took place on a larger register, the carry bit would propagate to the next $1$ and precipitate another carry.
After introducing a second factor of $p_\mathcal{N}$, the final probability of this second carry adder would be $p_\mathcal{N}^2$.
A perfectly error-free result can be obtained -- even with phase truncation -- if the partner bits of addition do not sum to the next level of significance.
In this sense, each rotation level can be perceived as a piece of required `memory' required to propagate the correct carry bits.

\subsection{Deriving an Expression For Probability Costs Incurred}
We seek a low-level expression that can be applied to provide the exact probability of obtaining the correct results under the addition of any two sized numbers for any truncation level.
Firstly, we begin with Equation (\ref{adder-expression}); this provides us a solid foundation for the perfectly working Draper adder, which can be modified to account for truncation.
In the case of truncation level $\mathcal{N}$, the sum in Equation (\ref{adder-expression}) is indexed from $k=j-\mathcal{N}$ to  $k=j$.
This omits the truncated rotations.
Consider the probability of the truncated adder given as
\begin{widetext}
\begin{equation}
\label{trunc-adder-expression}
    \begin{split}
        \text{Pr}\left(y_{L-1}\cdots y_0\right)&= |\langle y_{L-1}\cdots y_0|\psi\rangle|^2 \\
        & = \left| \frac{1}{2^L} \prod_{j=0}^{L-1}\left(\left(-1\right)^{y_j} + \exp\left[i\sum_{k=j-\mathcal{N}}^j 2\pi \left(\frac{x_k + a_k}{2^{j-k+1}}\right)-\sum_{k=j-\mathcal{N}+1}^{j}i\left(\frac{\pi y_{k-1}}{2^{j-k+1}}\right)\right]\right)\right|^2.
    \end{split}
\end{equation}
\end{widetext}
Without loss of generality, we consider the case where the first carry bit occurs at $k'=j-\mathcal{N}-1$.
Up until this, each qubit along the line is in a definite state of $\ket{0}$ or $\ket{1}$.
At the index $j = k'+\mathcal{N}+1$, we reach the case where the furthest $y_k$ reads an unaccounted-for $1$, and hence have a modified probability of
\begin{equation}
    \frac{1}{4}\left|1+\text{e}^{2\pi i\left(-\frac{y_{j-\mathcal{N}-1}}{2^{j-(j-\mathcal{N})+1}}\right)}\right|^2 = \frac{1}{4}\left|1+\text{e}^{-\frac{\pi i}{2^\mathcal{N}}}\right|^2.
\end{equation}
In the progressive evaluation each qubit before this point was in a definite state.
At the point where the carry bit origin is truncated, we now have $\ket{\psi} = \sqrt{p_\mathcal{N}}\ket{\text{correct}} + \sqrt{1-p_\mathcal{N}}\ket{\text{incorrect}}.$
Note that from here on we will primarily be concerned with the probability of obtaining a correct value, meaning being loose with square roots of absolute values. Once we end up in an `incorrect' state, it becomes exponentially unlikely to return to the correct one. 
Consider the next qubit along the line, $j = k' + \mathcal{N} + 2$.
Any further probability will be taken from the $\ket{\text{correct}}$ state, and so we can multiply out Equation (\ref{trunc-adder-expression}) for each individual qubit, supposing that the remaining qubits are in an exact state, and then take the final probability.
This next qubit will have probability of being correct: 
\begin{equation}
    \frac{1}{4}\left|\left((-1)^{y_j} + \text{e}^{\left[2\pi i\left(\frac{x_{j-\mathcal{N}}+a_{j-\mathcal{N}}}{2^{j-(j-\mathcal{N})+1}} - \frac{y_{j-\mathcal{N}}}{2^{j-(j-\mathcal{N})}}\right)\right]}\right)\right|^2,
\end{equation}
 which is equal to 
\begin{equation}
    \frac{1}{4}\left|\left((-1)^{y_{k'+\mathcal{N}+2}} + \text{e}^{\left[2\pi i\left(\frac{x_{k'+2}+a_{k'+2}}{2^{\mathcal{N}+1}} - \frac{y_{k'+2}}{2^{\mathcal{N}}}\right)\right]}\right)\right|^2.
\end{equation}
Under the supposition that we had a carry bit at position $k'$, then $y_{k'+2}$ will be a $1$ if and only if  $x_{k'+1} + a_{k'+1} \geq 1$.
That is, the error gets no worse if and only if $x_{k'+1}=a_{k'+1}=0$.
Otherwise we have $\ket{\psi}=\sqrt{p_\mathcal{N}^2}\ket{\text{correct}} + \sqrt{1-p_\mathcal{N}^2}\ket{\text{incorrect}}$.
From this point onward the process begins again. Once we leak some amplitude to the incorrect states, it never returns. 
The amplitude of the $\ket{\text{correct}}$ state is left the same if any $x_j + a_j < 2$, is multiplied out by $p_\mathcal{N}$ if another carry bit is encountered, and will continue to multiply out if that carry bit propagates.
When a carry bit begins this sequence of errors in $1s$ we refer to this as a \emph{carry chain}.
For the generic case of $x_j+a_j = 2$ we shall refer to as a carry \emph{event}. This allows us to simplify much of the previous calculations into the question: `when two numbers are added together, how many carry chains are there, and how long is each carry chain?'. This idea is depicted in Figure~\ref{fig:trunc-main}b. This then fully determines the error incurred by using the truncated Draper adder.
The total probability $\mathscr{T}$ of obtaining the correct result can be concisely expressed as
\begin{equation}
\label{trunc_expression}
    \mathscr{T} = \prod_{i=1}^{c}p_{\mathcal{N}}^{\min\{\left(y_i-\mathcal{N}\right)\cdot\Theta\left(y_i-\mathcal{N}\right),x_i\}}.
\end{equation}
Here, $c$ is the number of distinct carry chains; $\mathcal{N}$ is the level of truncation; $x_i$ is the length of carry chain of $1$s along which the carry bit propagates -- by this we mean the distance between where the carry bit started, and where it ends up; $y$ is the position of the left-most carry bit, counting up from 1; and $\Theta$ is the Heaviside step-function.
These components summarise the notion that $\mathscr{T}$ is given by the carry fidelity $p_\mathcal{N}$ as the base.
The power is the number of distinct carry bits multiplied, by the length of each of the respective carry chains, and confined to the first $L-\mathcal{N}-1$ qubits. \par
There is a symmetry here with subtraction through negative rotations.
The equivalent `memory' process in subtraction is that of borrowing.
When two bits align to $0-1$, a borrow bit must be taken from the next level of significance.
This borrowing will keep propagating along a chain of zeros until it encounters a $1$ at which point the chain will terminate.
The phase rotation will be in the opposite direction, but the magnitude of the error will be the same.

\subsection{Average error incurred}
Focusing now on the performance of a typical quantum algorithm, we average Equation (\ref{trunc-adder-expression}) over all $x$ and all $a$.
In particular, this is the case wherein $x$ truly unknown (initialised in equal superposition) until measurement, but $a$ will be known in the specific algorithm case. As mentioned, analysis here will be conducted as though $x$ and $a$ both have equal probability of 0 or 1 in each bit position.
That is, this will be the \emph{average} performance of the \emph{average} algorithm.
Application of this result to a particular $a$ will provide a more focused prediction for a given algorithm. The probability of colliding 1s could be accounted for in this updated case.
We treat two cases: first, an exact -- but computationally difficult -- expression in terms of $L$ and $\mathcal{N}$; and second, an asymptotic expression for large $L$.
We summarise our results here:
\begin{figure*}
    \centering
        \includegraphics[width=\linewidth]{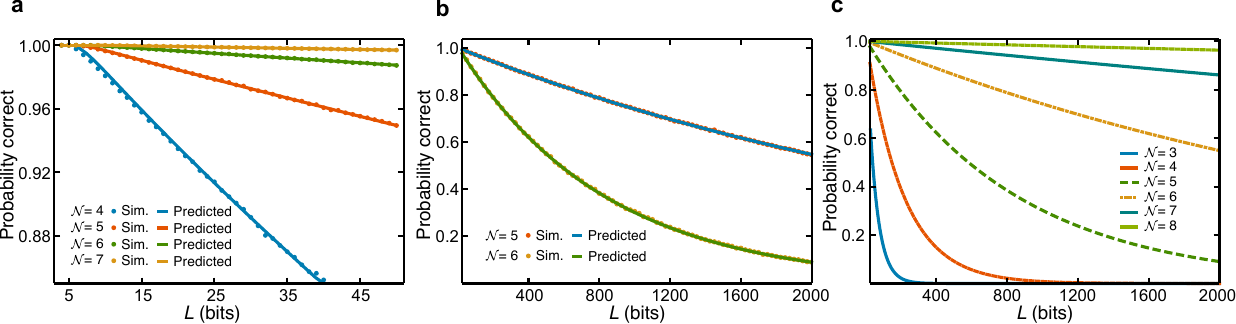}
        \caption{A comparison of adder predictions with simulation results. \textbf{a} Compares the small-$L$ regime with the full MPS simulations at 1000 instances. \textbf{b} compares the results of the asymptotic calculation to a Python simulation of a single state at 100 instances. \textbf{c} A prediction of the asymptotic behaviour of the truncated Fourier adder for different levels of $\mathcal{N}$}
        \label{adder_predictions}
\end{figure*}
\begin{theorem}
When two uniformly random numbers are summed together in an $L$-qubit Draper adder truncated to a level $\mathcal{N}$, the average probability of measuring the correct state is:
\begin{equation}
\label{trunc-L}
\begin{split}
        &\mathscr{T}_A(L,\mathcal{N})= \\
        &\left|\frac{1}{2}\left(\exp\left(-\frac{\pi i}{2^\mathcal{N}}\right)+1\right)\right|^{2\times C(L-\mathcal{N}-1)\times A\left(\tfrac{3}{4},L-\mathcal{N}-1\right)}.
\end{split}
\end{equation}
\text{Where:}
\begin{equation}
    A(p,n) := \sum_{k=0}^{n}\frac{S(p,n,k)\cdot k}{R(p,n)},
\end{equation}
\begin{equation}
    C(L) = \frac{L}{4\cdot \left(1 + \frac{1}{3}\cdot\left(A(\tfrac{3}{4},L) - 1\right)\right)},
\end{equation}

\begin{equation}
S(p,n,k):=\sum_{x=0}^{n} P_2(p,n,k,x)\cdot x,
\end{equation}
\begin{equation}
\begin{split}
    P_2(M_n^{(k)}=x) &:= P_2(p,n,k,x) \\
    &= P(p,n,k,x)-P(p,n,k+1,x),
    \end{split}
\end{equation}

\begin{equation}
    R(p,n) = \sum_{x=1}^{n}P(p,n,1,x)\cdot x,
\end{equation}
and,
\begin{equation}
\label{bernoulli-prob}
\begin{split}
    P(M_n^{(k)}=x)&:= P(p,n,k,x) \\
    &= \sum_{m=x}^{\lfloor\frac{n+1}{k+1}\rfloor} (-1)^{m-x}\binom{m}{x}p^{mk}q^{m-1}\times\\
    &\left(\binom{n-mk}{m-1}+q\binom{n-mk}{m}\right).
\end{split}
\end{equation}

\end{theorem}

We provide the full proof in Appendix \ref{sec:app-av}.
In short, we model the bits as two strings of Bernoulli random variables.
Here, the probability of being a 1 can be chosen if more information is known about $x$ or $a$.
We then start with the probability of a carry occurring (that is, the collision of a 1 in an $x$ bit matched by a $1$ in the $a$ bit).
We then find the average number and average distance of a carry chain, given that a carry bit occurring at the $(i-1)$th location is propagated by a 1 either on the $a_i$ or the $x_i$.\par
Equation (\ref{trunc-L}) contains many different interrelated quantities, but ultimately only depends on the average length of a carry chain. 

Therefore, we have:
\begin{corollary}
In the limit of large $L$, the average total probability is given by
\begin{equation}
\begin{split}
        \mathscr{T}_A(L,\mathcal{N})= \left|\frac{1}{2}\left(\exp\left(-\frac{i \pi}{2^\mathcal{N}}\right)+1\right)\right|^{L-\mathcal{N}-1}.
\end{split}
\label{asymptotic-prob-adder}
\end{equation}

\end{corollary}
\emph{Proof.}
For a large number of trials, the sum to compute the average length of a run tends towards infinity. 
Using the well-known result that
\begin{equation}
    \langle R(p)\rangle = \sum_{r \geq 1} r\cdot \text{P}(R = r) = \sum_{r\geq 1}\text{P}(R\geq r)
\end{equation}
and the fact that $\sum_{r\geq 1}\text{P}(R\geq r) =\sum_{s\geq 0}p^s$, then using the geometric series we have
\begin{equation}
    \langle R(p)\rangle = \frac{1}{1-p},
\end{equation}

where $p$ is the probability of propagation. \par 
For this model, $p=3/4$, and so the average length of a carry chain asymptotically tends towards 4. 
Applying Equation (\ref{trunc-L}) with an average length of 4, we have $1 + \frac{1}{3} \cdot (4 - 3) = 2$ carry bits per chain, giving a total of $\frac{L}{8}$ distinct carry chains, each of which has an average length of 4. 
The total number of errors therefore tends towards $(L-\mathcal{N}-1)/8\times 4$.\par

Equation (\ref{trunc_expression}) showed that the correct probability of any two numbers can be straightforwardly calculated without the need for a quantum simulation. 
A Monte Carlo simulation to compute this average fidelity of truncated addition was written in Python for large $L$ cases. 
Figure \ref{adder_predictions}b compares these results with Equation (\ref{asymptotic-prob-adder}). 
In addition, Figure \ref{adder_predictions}c demonstrates the predicted probability decay for different truncation levels with increasing $L$. This exceeds 50\% success probability for $\mathcal{N}\geq 6$. 


%% file: trunc_depth.tex
\section{Propagation of the Error with Depth}
\label{sec:trunc-depth}
\begin{figure}
    \centering
    \includegraphics[width=0.9\linewidth]{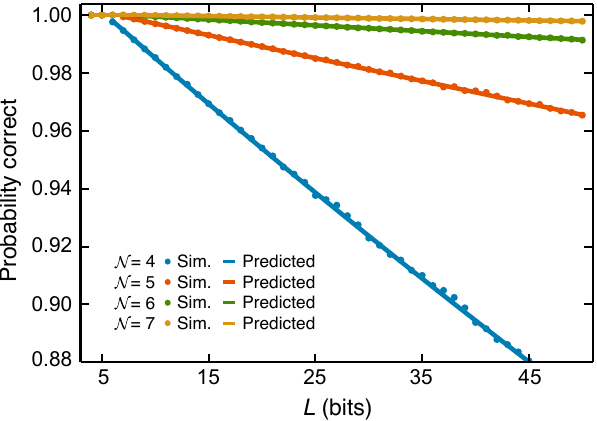}
    \caption{MPS simulations compared with analytic predictions for the average case of $x + a - b$ over a range of different $L$ and different $\mathcal{N}$.}
    \label{fig:as_predict}
\end{figure}

Equipped with accurate predictions about the behaviour of truncation in circuit size, we examine behaviour in depth. 
Addition finds its value by composing larger arithmetic operations through a series of repetitions. 
In this section we examine the depth scaling of the induced truncation errors. 
In general, these arithmetic circuits will be constructed through a series of additions \emph{and} subtractions. The primary reason for this is to ensure that average bit values remain at zero and errors cancel out. Repeated additions in the truncated regime will quickly fail (as carries become almost certain), but we mitigate this by performing addition by subtracting a number's 2's complement.
Prior to considering the behaviour of an arbitrary number of additions and subtractions, we consider the case of a single addition and a single subtraction. 
An adder and subsequent subtractor can be cast in a similar form to Equation (\ref{adder-expression}) with the further negative rotations included:
\begin{widetext}
\begin{equation}
\label{trunc-adder-subtractor-expression}
    \begin{split}
        \text{Pr}\left(y_{L-1}\cdots y_0\right)&= |\langle y_{L-1}\cdots y_0|\psi\rangle|^2 \\
        & = \left| \frac{1}{2^L} \prod_{j=0}^{L-1}\left(\left(-1\right)^{y_j} + \exp\left[i\sum_{k=j-\mathcal{N}}^j 2\pi \left(\frac{x_k + a_k - b_k}{2^{j-k+1}}\right)-\sum_{k=j-\mathcal{N}+1}^{j}\left(\frac{\pi y_{k-1}}{2^{j-k+1}}\right)\right]\right)\right|^2
    \end{split}
\end{equation}
\end{widetext}
Equation (\ref{trunc-adder-expression}) showed that from the IQFT emerged errors through the net sum of rotations on a given qubit.
For this reason, a carry bit can be cancelled out by a subtraction on the same qubit. 
A truncation effect is consequently induced with the presence of either a net carry bit $1+1-0$ or a net borrow bit $0+0-1$. 
The probability of error is therefore $\frac{1}{8} + \frac{1}{8} = \frac{1}{4}$, the same as the adder case on its own. 
A difference arises in the probability of propagation of an error chain. 
Beginning with a carry bit, the chain can be halted by a $0+0-0$, $1+0-1$, $0+1-1$ or $0+0-1$. 
Similarly, the chain from a borrow bit is interrupted by a $0+1-0$, $1+0-0$, $1+1-1$, or $1+1-0$; propagation now only $1/2$ as likely to occur, rather than $3/4$. 
Applying the same tools as with Equation (\ref{trunc-L}), the average number of distinct chains is given by:
\begin{equation}
    B(L)=\frac{L}{4\cdot \left(1 + \frac{1}{2}\cdot\left(A(\tfrac{1}{2},L) - 1\right)\right)}.
\end{equation}
Once more, this gives the overall expected fidelity $\mathscr{T}_{AS}$ at a truncation level $\mathcal{N}$ and qubit size $L$:
\begin{equation}
\begin{split}
        &\mathscr{T}_{AS}(L,\mathcal{N}) = p_{\mathcal{N}}^{ B(L-\mathcal{N}-1)\times A(\tfrac{1}{2},L-\mathcal{N}-1)}.
\end{split}
\end{equation}
Figure \ref{fig:as_predict} compares this analytic model with MPS simulation results over a range of $L$, showing good agreement.

Similarly, the asymptotic case can be evaluated. 
\begin{corollary}
The expected length of a given success run is $\langle R \rangle = \frac{1}{1-\frac{1}{2}} = 2$. 
The total number of errors is therefore $(L-\mathcal{N}-1)/6\times 2$, giving the final probability of a correct result as:
\begin{equation}
\begin{split}
         \mathscr{T}_{AS}(L,\mathcal{N})= p_{\mathcal{N}}^{\frac{(L-\mathcal{N}-1)}{3}}.
\end{split}
\label{AS-average-asymptotic}
\end{equation}
\end{corollary}
When making a comparison between Equations (\ref{AS-average-asymptotic}) and (\ref{asymptotic-prob-adder}), it is clear that the performance of an adder followed by a subtractor is better than an adder alone, despite requiring double the operations. 
This natural method of cancelling out errors will form one of the key components of how we construct our arithmetic operations using the truncated adder.

\subsection{Truncation Scaling with Repeated Additions}
The previous sections provided an understanding of the scaling behaviour of a truncated adder in size. 
In order to construct larger arithmetic operations, repetitions of an adder are necessary. 
For this reason, it is important to ascertain the modulation of behaviour in depth as well as size. 
In typical error analyses, an erroneous component probability multiplies out. 
However, given that the errors occur in the truncated Fourier basis, the behaviour is not this simple. \par
Larger operations such as multiplication can be constructed using only a sequence of adders. 
However, the previous section made clear that a subtraction of another number helped to suppress the truncation errors. 
For this reason, instead of only using repeated adders, we consider an alternating series of adders, followed by subtractions of a number's \emph{two's complement}. 
The two's complement is defined for a number $a$ with binary length $L$ as being $2^L - a$. 
Since binary addition and subtraction is modulo $2^L$, then instead of computing $a+b$, the calculation can be $a - (2^L - b)\:\text{mod}\:2^L$ = $a+b$. 
In full generality then, we aim to compute the total correct probability with a sequence of $n_1$ adders, and $n_2$ subtractors. 
This is equivalent to $n_1+n_2$ individual adders. 
The maximum likelihood of error cancellation is when $n_1=n_2$. 
Given that addition and subtraction are equally difficult to perform, we will operate under this assumption. 
However, a circuit with information about the structure of numbers could modify $n_1$ and $n_2$ in order to produce the highest probability of success.\par
The problem is set up as follows: an $L$-qubit quantum register with initial value $x$ undergoes $n$ additions and $n$ subtractions. 
This is represented by $y = x + a_1 - b_1 + a_2 - b_2 + \cdots + a_n - b_n$. 
We will denote $y_i$ to be the $i^{\text{th}}$ bit of the final outcome, and $c_i$ to be the total vertical sum in the $i^{\text{th}}$ position. 
That is, $c_i = x_{i} + a_{1,i} - b_{1,i} + a_{2,i} - b_{2,i} +\cdots a_{n,i}-b_{n,i} = x_i + \sum_{k=1}^{n_1}a_{k,i} - \sum_{k=1}^{n_2} b_{k,i}$. 
Equation (\ref{adder-expression}) can be generalised to introduce each of the $2n$ phase rotations as follows:
\begin{widetext}
\begin{equation}
\label{multiple-adder-expression}
    \begin{split}
        \text{Pr}\left(y_{L-1}\cdots y_0\right)&= |\langle y_{L-1}\cdots y_0|\psi\rangle|^2\\
        & = \left| \frac{1}{2^L} \prod_{j=0}^{L-1}\left(\left(-1\right)^{y_j} + \exp\left[i\sum_{k=0}^j 2\pi \left(\frac{x_k + \sum_{i=1}^{n_1} a_{i,k} - \sum_{i=1}^{n_2} b_{i,k}}{2^{j-k+1}}\right)-\sum_{k=1}^{j}\left(\frac{\pi y_{k-1}}{2^{j-k+1}}\right)\right]\right)\right|^2.
    \end{split}
\end{equation}
\end{widetext}
The index $i$ rotations constitute all additive and subtractive rotations in the $i$th binary position, and the product is the whole horizontal span. 
This will be referred to as a grid, where each row constitutes a binary representation of a particular number, and each column isolates the net sum of a given bit. 
Moving to the right in a row is less significant, and conversely moving to the left is more significant. 
The immediate problem is that the over or under rotations can now be extended out far beyond a single bit. 
The probability, and contributing magnitude of these effects must all be computed.
\subsubsection{Quantifying the Contribution of Multiple Carries on a Single Qubit}
When a carry event occurred on a truncated bit, the resulting IQFT caused an under-rotation of $-\pi/2^{\mathcal{N}}$. 
This effect generalises; whenever a truncated bit $j$ sums to a number greater than 1, its effects will travel along in the final result. 
Consequently, its influence will be found in the IQFT phase rotation on the $(j+\mathcal{N}+1)^\text{th}$ qubit. 
The sum will give the amount by which the IQFT under-rotates. 
This is quantified as follows: consider a scenario with truncation level $\mathcal{N}$. 
The $j$th qubit can sum to $a_n\cdots a_0$, for some length $n$. 
We denote this sum as $c_j$. The probability that rests on the $(j+\mathcal{N}+1)$th qubit will be
\begin{equation}
\begin{split}
       &\frac{1}{4}\left|1+\exp\left[2\pi i\left( - .a_n\cdots a_1\right)\right]\right|^2 \\
        &= \frac{1}{4}\left|1+\exp\left[i \pi\left(-\left\lfloor \frac{c_j}{2}\right\rfloor \frac{1}{2^\mathcal{N}}\right)\right]\right|^2,
\end{split}
\end{equation}
where the equality to the floor of $c_j$ follows since the truncation will omit the $a_0$ part of $c_j$. \par
In the same way that the single adder did not reduce to finding the probability of $1+1=2$, the challenge here is not only to isolate the distribution of $c_j$. 
In particular, if the sum of lesser significant bits $c_i$ are greater than 2, they will impact $c_{i+1}$. 
This is an extension of the carry-chain idea, wherein carry bits propagated through different occurrences of a $1$.
This needs to be taken into account when predicting all of the effects of truncation. 
The effects of the lesser significant bits must be accordingly weighted, in order to account for the possibility of them adding up into something \emph{just as} significant. 
For every position along a row, the bit to the right are weighted 1/2 as much, and then 1/4 as much, and so on. We define, therefore, a final variable of interest $d_j$ which we say is the \emph{effective} sum in the $j$th level of significance. 
We define it by:
$$d_j := \sum_{k=0}^j \frac{1}{2^{j-k}}c_k$$
For every $d_j$, the probability factor for obtaining the correct result on the $(j+\mathcal{N}+1)$th qubit is therefore:
\begin{equation}
    \frac{1}{4}\left|1+\exp\left[i \pi\left(-\frac{\lfloor d_j/2 \rfloor }{2^{\mathcal{N}}}\right)\right]\right|^2
\end{equation}
The \emph{total} probability of obtaining the correct result is consequently:
\begin{equation}
    \frac{1}{2^{2L}} \prod_{j=0}^{L-\mathcal{N}-1} \left|1+\exp\left[i \pi\left(-\frac{\lfloor d_j/2\rfloor }{2^{\mathcal{N}}}\right)\right]\right|^2
\end{equation}
This is also true in the case of subtraction. 
Previously, it was shown how a borrow bit would result in an over-rotation instead of an under-rotation; where the information is taken from the left of the bit string rather than the right. 
The two cases are entirely symmetrical. \par
Summarising the problem, therefore, we must determine the distribution of $L-\mathcal{N}-1$ (not independent) $d_j$ random variables, where each $d_j = \sum_{k=0}^j \frac{1}{2^{j-k}}c_k$, each $c_k = x_k + \sum_{i=1}^{n_1}a_{i,k} - \sum_{i=1}^{n_2} b_{i,k}$, and each $a_{i,j}$ and $b_{i,j}$ can be either 0 or 1 with a given probability. 
A characterisation of this will entirely determine the behaviour of the truncated adder.

\subsubsection{Distribution of Repetition Errors}
In the previous section, we were able to calculate exactly the distribution of truncation errors even for small numbers. 
For this section, we will focus only on the asymptotic case. 
The reason is that our concept of an `error chain' is no longer binary, in the sense that it now exists in different magnitudes depending on the value of each $d_j$. 
The smaller the $L$, the more conditional the errors are on their surrounding qubits. 
With the total combinations increasing factorially, if a closed form of the nested conditional probability exists, it is likely not simple. 
Instead, we will work on the asymptotic case, compare our conclusions to the small $L$ simulations, and compare how the two differ. 
In the asymptotic case, the average qubit error is insensitive of the surroundings. 
The proportion of qubits with a correct probability of $\frac{1}{4}\left|1+\exp\left[i \pi\left(-\frac{\lfloor d_j/2 \rfloor }{2^{\mathcal{N}}}\right)\right]\right|^2$ is exactly $\text{Pr}\left(d_j \leq D_j < d_j + 2\right)$, where $D_j$ is the random variable of the \emph{value} of $d_j$. 
We will now compute the probability distribution of $D_j$. 
Recall that the random variable $c_j = x_j + \sum_{k=0}^{n_1} a_{j,k} - \sum_{k=0}^{n_2} b_{j,k}$. 
This simplifies as the difference of two binomially distributed\footnote{The notation $\stackrel{d}{=}$ is used to denote `sampled from this distribution'} variables $A - B$, where $A\stackrel{d}{=}\text{Bi}(n_1+1,\frac{1}{2})$ and $B\stackrel{d}{=}\text{Bi}(n_2,\frac{1}{2})$. 
In general, with $A \stackrel{d}{=} \text{Bi}(n_1,p_1),\: B \stackrel{d}{=} \text{Bi}(n_2,p_2)$, then the support of $C = A - B$ is $[-n_2, n_1]$. 
We need to count up all the ways in which we can have $c = a - b$ for some given $c$. 
The case of $c\geq 0$ and $c<0$ are treated separately. 
For $c\geq0$, $c$ can be obtained with $a = i + c$ and $b = i$, for some $i$ in the range of $A$. 
The probability of obtaining $c$ is then $\text{Pr}(A=i+c)\cdot\text{Pr}(B=i)$, summed over all $i$. 
Since $A$ and $B$ are binomially distributed, they have the usual PMF of $$\text{Pr}(X=k) = f(k;n,p)=\binom{n}{k} p^k (1-p)^{n-k};\:k\leq n,$$ and 0 otherwise. 
Overall this can be summarised as:
$$\text{Pr}(C = c) = \sum_{i=0}^{n_1} f(i+c; n_1, p_1)\cdot f(i;n_2,p_2)$$
Similarly, the case of $c<0$ has the roles reversed. 
This gives us the overall PMF of:
\begin{equation}
\label{bin-difference}
\text{Pr}(C=c) =
    \begin{cases}
      \sum_{i=0}^{n_1} f(i+c; n_1, p_1)\cdot f(i;n_2,p_2) & c\geq 0 \\
      \sum_{i=0}^{n_2} f(i; n_1, p_1)\cdot f(i-c;n_2,p_2) & c<0 \\
   \end{cases}
\end{equation}
The variable $D_j$ is given by a scaled sum of the $C_j$. 
The distribution must firstly be rescaled: $\text{Pr}(k\cdot C = c) = \text{Pr}(C = c/k)$ if $c/k$ is an integer, and $0$ otherwise. 
We designate $\mathcal{P}_{j,k}$ for the PMF of $C_i$ scaled by a factor $k$. 
The PMF of a variable which is the sum of other random variables is given by the discrete convolution $\ast$ of the individual PMFs in the summand. 
Consequently, the PMF of $D_i$ -- which we designate $\mathscr{P}_j$ is:
\begin{equation}
\label{general-pmf}
    \mathscr{P}_j=\mathop{\scalebox{2.5}{\raisebox{-0.2ex}{$\ast$}}}_{k=0}^i \mathcal{P}_{k,\frac{1}{2^{i-k}}},
\end{equation}
where the notation used denotes an $n-$fold convolution, as described above.
The support of a single $C_{j,k}$ is $[-\frac{n}{k}, \frac{n+1}{k}]$. 
We can efficiently compute $\mathscr{P}_i$ by first calculating the full probability mass function over the support of each $C_{j,k}$, giving us a list of probabilities. 
We then perform a fast Fourier transform on each list. 
Once in Fourier space, the convolution of two functions can be performed by multiplying them together, and taking the inverse Fourier transform.\par
In the effective sum comprising $D_j$, we have exponentially diminishing contributions from each value to the right. 
It is therefore unnecessary to account for all lesser significant bits. 
\begin{figure}[ht]
    \centering
    \includegraphics[width=0.9\linewidth]{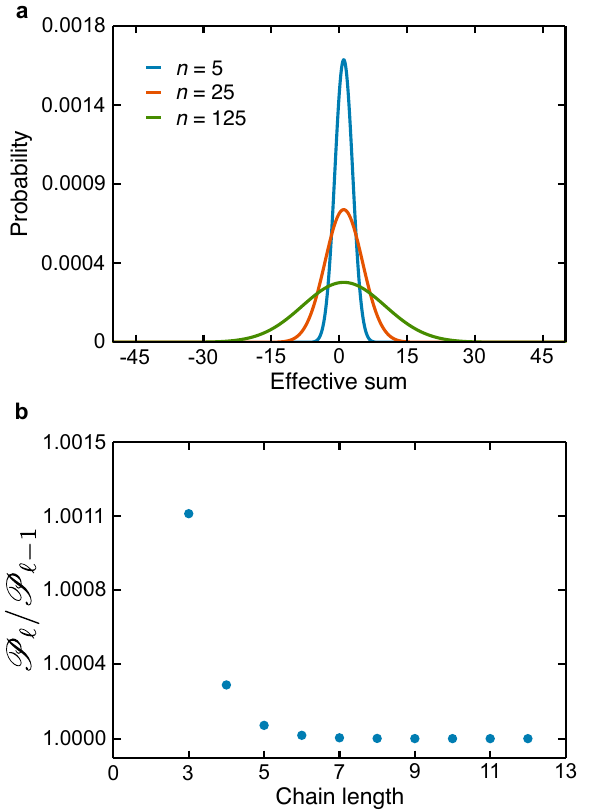}
    \caption{Example summary of computed PMFs from Equation \eqref{general-pmf}. \textbf{a} Illustration of the widening of the probability distributions with increasing sequences of additions and subtractions; the larger the absolute value of the effective sum, the more damaging the error is. \textbf{b} Accounting for larger chain lengths in the probability computation for $n=500$, we find exponentially diminishing contributions for each extra included bit. Because of the quick convergence to 1, we take $\ell=8$ in our calculations.}
    \label{fig:pmf_summary}
\end{figure}
We take a chain of length $\ell$, where $\ell$ is sufficiently large to capture all significant probabilities of error (we will soon explore what `sufficiently large' means in this context). 
This greatly simplifies computation. Moreover, because this is the asymptotic case, it means we can apply the calculation to all the bits in the string. 
That is $D_j \equiv D$, where, $D$ now represents \emph{any} bit along the line. 
Let us now denote $D(n)$ as the random variable $D$ after $n$ additions and $n$ subtractions. 
Figure \ref{fig:pmf_summary}a, shows the probability distribution of $D(n)$ for different values of $n$. 
As might be expected, the probability mass function looks like an interpolated binomial distribution. 
These PMFs were constructed with a $D$ using a chain-length of 8. 
Using $n=500$ as a case-study, the probability of a carry error was computed for a range of $\ell$, and then the ratio with the previous $\ell$ calculated. 
These results are shown in Figure \ref{fig:pmf_summary}b and demonstrate the speed with which the PMF converges with $\ell$, verifying our choice. 
It is evident that with an increasing $n$ we have an increasing variance. 
This is what leads to a more damaging truncation effect with sequential adders.\par
The variance of the sum of $n$ binomially distributed variables is $np(1-p) = n/4$. 
In our case, each $C_j$ is the sum of $2n+1$ binomial variables. 
As such, it is distributed with variance $(2n+1)/4$. Next, we note that $\text{Var}[k\cdot X] = k^2 \text{Var}[X]$ for any random variable $X$ and any constant $k$. 
Hence:
\begin{equation}
\label{variances}
    \begin{split}
    \text{Var}[D] &= \sum_{k}\text{Var}\left[\frac{C_k}{2^k}\right]\\
    &= \sum_k \frac{1}{2^{2k}} \text{Var}\left[C_k\right]\\
    &= \left[\sum_k\left(\frac{1}{4}\right)^k\right]\frac{2n+1}{4}\\
    &= \frac{2n+1}{3}
    \end{split}
\end{equation}
With $\mathscr{P}(D)$ well-characterised, the performance of sequential additions and subtractions can be evaluated. 
The support of the random variable $D$ is given by the sum of each $C_j$. 
That is, the lower bound is $\sum_{k}\left(\frac{1}{2}\right)^k \cdot (-n) = -2n$ and the upper bound is $\sum_{k}\left(\frac{1}{2}\right)^k \cdot (n+1) = 2(n+1)$. 
The support here represents the spectrum of possible errors. 
In an extension of our earlier use of a carry fidelity $p_\mathcal{N}$ we define a class of carry fidelities: $p_{\mathcal{N},a} := \left|\frac{1}{2}+ \frac{1}{2}\text{e}^{-\frac{i \cdot a\cdot \pi}{2^\mathcal{N}}}\right|^2$, where $a$ is an integer. 
Then, the total probability of obtaining the correct result (with the product taken over all non-integer $d$) is:
\begin{equation}
\label{probability-repeated-first}
    \prod_{d=-2n}^{2n+2} p_{\mathcal{N},\lfloor d/2\rfloor}^{\text{Pr}\left(D = d\right)\cdot L}.
\end{equation}
This expression is messy, and at its surface divulges very little superficial behaviour about the propagation of the truncated error. 
The reason for this is twofold: the size of each $p_{\mathcal{N},a}$ with the scaling of $a$ is unclear; and the sequential probabilities require intensive pre-calculating. 
We now aim to relate the general fidelity $p_{\mathcal{N},a}$ to our original carry fidelity. 
Suppose we take $p_{\mathcal{N},a}$. 
The argument of the exponential in this is typically very small. 
Applying the small-angle approximation yields the following relationship:
\begin{equation}
    \begin{split}
        p_{\mathcal{N},a} &= \frac{1}{4}\left|1 + \text{e}^{\left(\frac{a\cdot \pi \cdot i}{2^\mathcal{N}}\right)}\right|^2 \\
        &= \frac{1}{4} \left(2 + 2\cos\left(\frac{a\cdot \pi \cdot i}{2^\mathcal{N}}\right) \right),\\
        &\approx \left(1 + \frac{1}{4} \left(\frac{\pi}{2^\mathcal{N}}\right)^2\right)^{a^2},\\
        &\approx \left(\frac{1}{4}\left(2 + 2\cos\frac{\pi}{2^\mathcal{N}}\right)\right)^{a^2}= p_{\mathcal{N},1}^{a^2}.
    \end{split}
\end{equation}
Using this approximate relationship, we can re-express Equation (\ref{probability-repeated-first}) as
\begin{equation}
\label{as-correct-series}
   |\langle\psi|\text{correct}\rangle|^2 = p_{\mathcal{N},1}^{\sum_{d=-2n}^{2n+2}(\lfloor d/2 \rfloor)^2\cdot \text{Pr}\left(D = d\right)\cdot L}.
\end{equation}

We are now ready to derive a simple expression for the probability of obtaining the correct result in repeated truncated addition and subtraction.
\begin{theorem}
In the case of truncation to a level of $\mathcal{N}$ for a binary number of length $L$ with $n$ additions and $n$ subtractions, the probability $\mathscr{T}_{AS}(n)$ of obtaining the correct result is given asymptotically by:
\begin{equation}
\label{repeated-adder-probability}
    \mathscr{T}_{AS}(n) = \left|\frac{1}{2}+\frac{1}{2}\exp\frac{i \pi}{2^\mathcal{N}}\right|^{2L\left(\frac{n+1}{6}\right)}.
\end{equation}

\end{theorem}

We provide the full proof of this in Appendix \ref{sec:app-av-depth}.

The appearance of the $d^2$ in this series is completely unrelated to the probabilities themselves, and the probabilities are difficult to evaluate -- so it is somewhat surprising that this all arrives at a result in which we have a single base and a single power which is linear in $n$. 
It is fortunate, however, that we can summarise the behaviour of this complex system in a single digestible equation. 

\begin{figure}[ht!]
    \centering
        \includegraphics[width=\linewidth]{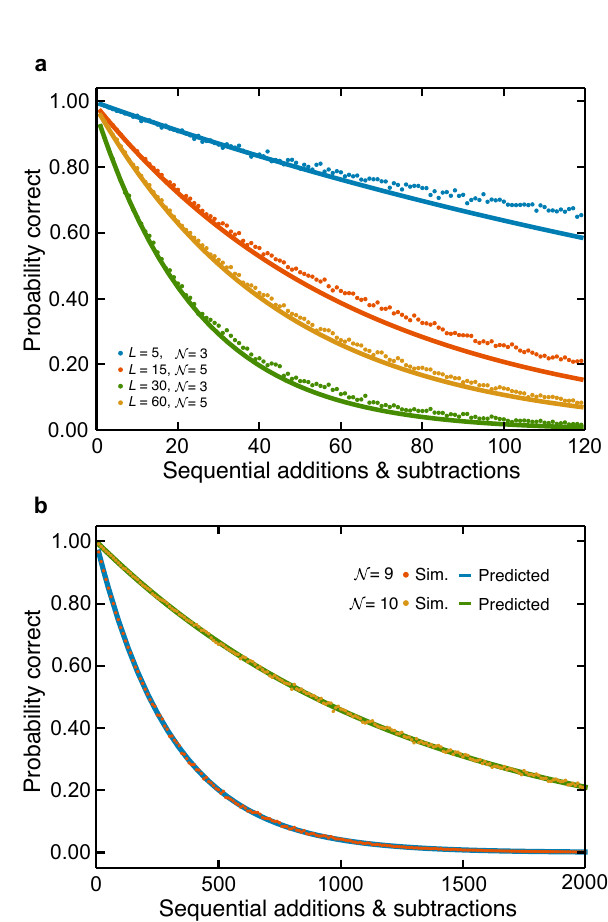}
        \caption{A comparison of model predictions with simulation results. \textbf{a} shows that in the low $L$ regime, the $n/2$ power decay does not entirely explain the behaviour of sequential additions. This is because our assumption of independence in the qubits is invalid for low $L$. With increasing $L$, the model becomes a better approximation. \textbf{b} in the large $L$ regime ($L = 2048$), the results appear to match up with predictions more precisely.}
        \label{sequentials}
\end{figure}

With a prediction model constructed, comparisons to simulation results can be made. 
To this effect, a Python simulation to compute Equation (\ref{multiple-adder-expression}) was developed. 
The results, for a Monte-Carlo simulation of 2048 bit numbers are shown in Figure \ref{sequentials}b, with Equation (\ref{repeated-adder-probability}) overlaid on top. 
In the asymptotic case we see a much better agreement of theory and simulation results. 
Both theory and data show a power decay which multiplies out on average every 12 adders or subtractors. 
For small $L$, the error situated on each qubit is correlated to its neighbours in a way that we have not accounted for in our calculations. 
Since our simulations of components of Shor's algorithm can only be conducted in the realm of relatively small $L$, we wish to visualise exactly by how much the above prediction of exponential decay differs from the actual simulation results.\par
With an exact result for the case of a single addition and subtraction, we can eliminate some error by recasting Equation (\ref{repeated-adder-probability}) as
\begin{equation}
\label{repeated-adder-simple}
    \mathscr{T}_{AS}(n) = \mathscr{T}_{AS}(1)^{\frac{n+1}{2}}.
\end{equation}
That is, we relate its growth to the application of a single adder and subtractor -- which could be numerically based -- rather than the carry fidelity. 
Figure \ref{sequentials}a shows this curve overlaid on top of MPS simulation results for $L=5,\:\mathcal{N}=3;\:L=15,\:\mathcal{N}=4;\:L=30,\:\mathcal{N}=4;\:L=60,\:\mathcal{N}=5$.
As $L$ grows, the predicted scaling matches simulation results better and better, as $D_j\rightarrow D$. 
We also see that the early behaviour with small $n$ matches that of prediction still quite well, and that we could use this model to at least predict to a good degree the probability of obtaining the correct result.
\subsection{Truncated Fourier Adder Summary}
We have shown that the effects of modifying a the QFT for arithmetic can be abstracted into an extensive evaluation of the distribution of $1$s and $0$s. 
This is our main contribution to the study of the AQFT; that its applications to quantum arithmetic are characterised by the input, not the operation itself. 
This result greatly modifies current savings estimates of Fourier-based arithmetic. 
The commonly taken level of truncation of the QFT in the literature is $\mathcal{N}=\log_2 \frac{L}{\epsilon}$ for some error $\epsilon$ \cite{coppersmith-est}. 
Typically, this error is expected to multiply out with repeated applications of the QFT. 
We have discussed how truncation in the context of Fourier addition scales for a given addition, or sequences of additions and subtractions. 
We can invert Equation (\ref{repeated-adder-probability}) to provide an expression for $\mathcal{N}$ for a given error $\epsilon$ and a given number of arithmetic operations $n$. 
To allow direct comparison to estimates of a multiplying error, $n$ refers to each application of either an adder or a subtractor. 
Equivalently, this is $n/2$ adders and $n/2$ subtractors -- hence $n=\mathcal{O}(1)$ for addition, $\mathcal{O}(L/2)$ for multiplication, and $\mathcal{O}(L^2/4)$ for exponentiation. 
This expression is:
\begin{equation}
\begin{split}
   1-\epsilon &= \left|\frac{1}{4}\left(1+\text{e}^{\frac{i\pi}{2^\mathcal{N}}}\right)\right|^{2\frac{n/2+1}{6}L}\\ &=\left[\frac{1}{2}+\frac{1}{2}\cos\frac{\pi}{2^\mathcal{N}}\right]^{\frac{n+2}{12}L},\\
    \Rightarrow \mathcal{N}&=\log_2\left(\pi\left(\arccos\left(4\left(1-\epsilon\right)^{\frac{12}{(n+2)L}}\right)-2\right)^{-1}\right).
\end{split}
\end{equation}
In general, this truncation provides a significant resource saving. Compared to $\mathcal{N} = \log_2\frac{L}{\epsilon}$, we find that the truncation level can be reduced to approximately half of this.
This corresponds to a saving of $\mathcal{O}\left(L\log_2\frac{L}{\epsilon}\right)$ rotation gates when compared to the current approximated values.


%% file: error_model.tex
\section{Modelling stochastic control errors in Fourier arithmetic}
\label{sec:error-model}
A large-scale quantum computer will possess resource requirements strongly related to the target error rate to which a quantum algorithm can withstand. For this reason, it is important to have a precise understanding of an algorithm's tolerance to error, such that their physical demands can be accurately evaluated and tailored appropriately.
We divert now to an assessment of the effects of rotation errors on the performance of the Fourier adder. Partly, this is because obtaining an accurate assessment of tolerance to error in Fourier arithmetic is crucial to estimating its resource requirements in the surface code, such as in Ref.~\cite{gidney2021factor}.
Previous estimates in the literature often evaluate this precision to first order at $\approx 1/n_p$, where $n_p$ represents the number of locations in which an error can occur. They are often limited to instances applicable to single circuit sizes \cite{miquel-1996}, focus solely on the period-finding subroutine \cite{lnn-robustness, chuang1995chuang}, or do not weight the difficulty of implementing different gates.\par 
\begin{figure*}[ht]
    \centering
    \includegraphics[width=\linewidth]{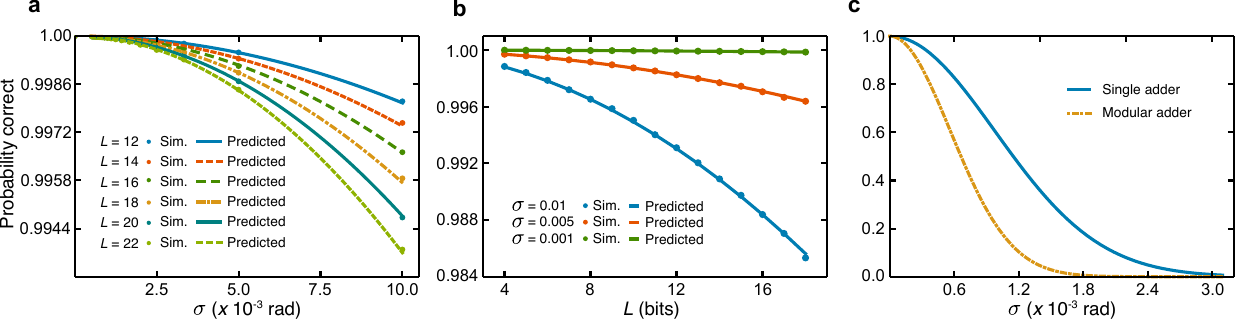}
    \caption{A comparison of imperfect rotation simulation results with an analytic model. \textbf{a} illustrates the behaviour of a single erroneous adder over a range of values for $L$ and $\sigma$. \textbf{b} demonstrates the same information for an isolated modular adder. \textbf{c} Extrapolates this circuit behaviour to large $L$}
    \label{errorgraphs}
\end{figure*} 
Here, we first investigate the robustness of the Fourier addition under the assumption of no truncation, but imperfect rotation gate fidelities. We look at the adder in the main text, and expand in Appendix \ref{app:shor-error} in the context of components of Shor's algorithm, followed by an evaluation of the entire circuit. Once the expressions of probability are derived for each component, we will use these to predict the circuit robustness when used in the case of $L=2048$. The phase gates in the QFT and adder of the Shor circuit are the only non-Clifford gates we encounter, and so following the magic states model we consider all rotations of $\leq\pi/4$ to have some inherent error after distillation, with all others to be perfect in their implementation.\par
With each operational implementation of a $Z-$rotation gate, it is likely that the true outcome is some fluctuation about the desired angle. Our model assumes that with each instance of a $Z-$rotation gate, $R_\phi$ performs the phase rotation $R_\phi \ket{1} \mapsto \text{e}^{i\left(\phi + \epsilon\right)}\ket{1}$ where $\epsilon$ is a random variable sampled from a Gaussian distribution with mean $\mu = 0$ and standard deviation $\sigma$. This allows for rotations which both exceed, and fall short of the target. The choice of $\mu=0$ is in keeping with full generality, where no systematic errors are expected. If these were observed to exist in a given architecture, proceeding calculations could be modified appropriately. Furthermore, when multiple errors accumulate in the phase of a given qubit state, they will sum together like $\exp\left(i\sum_k \epsilon_k\right)$. From the central limit theorem, the average sum of these random variables will quickly approach a Gaussian. For this reason, the total behaviour of the circuit will be largely insensitive to the parent distribution of $\epsilon$. Assuming the parent distribution to be Normal, therefore, is an assumption we expect can be made without consequence. \par 
Many noise models quantify their error rates through some measure that compares the distance of the ideal state density matrix with a state affected by the noisy channel \cite{Nielsen2010QuantumInformationb}. That is, a channel $\mathcal{E}$ with $Z$ noise performing an ideal operation $U$ transforms the density matrix $\rho$ into the state 
$$\mathcal{E}(\rho) = (1-\eta)U\rho U^\dagger + \eta Z U\rho U^\dagger Z^\dagger,$$
has an error rate $\eta$. In particular, this convention is used in \cite{rotation-magic-states} to categorise the fidelity of their distilled magic states. In Appendix \ref{app:error-ensemble}, we show that we can interpret this error model as being a $q=\frac{1}{2}\left(1-\text{e}^{-\frac{\sigma^2}{2}}\right)$ probability of a $Z$ flip on our qubit. $q$ here is exactly the $\eta$ from above. This gives an equivalence between our model of control errors and a phase-damping channel: this will be equally applicable in the case of both physical origins. This expression will be particularly relevant when the distillation cost of different gates is considered. For simplicity, we take the noise to be diagonal. This allows an analytic model for the effects of rotation error to be derived. In \cite{rotation-magic-states}, it is shown that distilled magic states suppress non-diagonal noise. We assume, then, that the effects of this are no worse than fluctuations around the $Z-$axis.\par 

The effects of this error are very similar to truncation effects. They will be present in each \textsc{C-Phase} gate in the QFT and IQFT, as well as each rotation gate in the adder itself. Equation (\ref{adder-expression}) can be modified to include the presence of these fluctuations, as well as the associated values of their controls. This yields a probability of:
\begin{widetext}
\begin{equation}
    \label{error-expression}
    \begin{split}
        \left| \frac{1}{2^L} \prod_{j=0}^{L-1}\left(\left(-1\right)^{y_j} + \exp\left[i\sum_{k=0}^j \left(\epsilon_{Q\:j-1,k}x_{k-1} + 2\pi \left(\frac{x_k + a_k}{2^{j-k+1}}\right)-\frac{\pi y_{k-1}}{2^{j-k+1}} + \epsilon_{I\:j-1,k}y_{k-1}\right)+i\epsilon_{P\:j}\right]\right)\right|^2.
    \end{split}
\end{equation}
\end{widetext}
This expression accounts for all of the phase rotations performed on any given qubit. In order to derive a stochastic model for the performance of the adder, we look to determining the average of \ref{error-expression}. Consider that, in general, only half of the controlled errors will in general be induced (only half of the $x_i$ and $y_i$ will be $1$), that all $\epsilon$ are sampled from the same distribution, and finally that $x_k + a_k - y_{k-1}$ will be either $0$ or $1$, then Expression (\ref{error-expression}) will look like
\begin{equation}
\label{factored-error-expression}
    \begin{split}
        \left|\frac{1}{2^L}\left(1+\text{e}^{i\epsilon_{0,0}}\right)\left(1+\text{e}^{i(\epsilon_{1,0}+\epsilon_{1,1})}\right)\cdots \left(1+\text{e}^{i\sum_{k=0}^j\epsilon_{j,k}}\right)\right|^2.\\
    \end{split}
\end{equation}

From here, an exact result can be derived for the average performance of an adder. A well-known result from probability theory is that the sum of two independent Gaussian-distributed random variables $x$ and $y$ is itself a Gaussian with variance the sum of the two individual variances, i.e., sampled from $N(0,\sigma_x^2 + \sigma_y^2)$. Since all the errors introduced in the circuit are distributed in the same way, we can simplify Equation (\ref{factored-error-expression}) by collecting together the $\epsilon_{j,k}$ in each exponential:
\begin{equation}
    \left|\frac{1}{2^L}\left(1+\text{e}^{i\epsilon_{1}}\right)\left(1+\text{e}^{i\epsilon_{2}}\right)\cdots \left(1+\text{e}^{\epsilon_{j}}\right)\right|^2,
\end{equation}
where $\epsilon_j \stackrel{d}{=} N(0,j\cdot\sigma^2)$. Each $\epsilon_j$ is an independent variable over which we can integrate. Since it remains confined to its own factor in the factorised expression, we can separate out the integrals. Then the exact average expression for the probability of obtaining the correct result with errors normally distributed with standard deviation $\sigma$ is:
\begin{equation}
    \label{exact-error}
    \begin{split}
       \langle P \rangle &= \int \cdots \int \text{d}\epsilon_1\cdots\text{d}\epsilon_L\\ &\left|\frac{1}{2^L}\left(1+e^{i\epsilon_{1}}\right)\cdots \left(1+\text{e}^{\epsilon_{j}}\right)\right|^2 \cdot P(\epsilon_1)\cdots P(\epsilon_L),\\
       &= \prod_{k=1}^L \int \text{d}\epsilon_k \frac{1}{4}\left(1 + \text{e}^{i\epsilon_k}\right)\left(1+e^{-i\epsilon_k}\right)\cdot  \frac{\text{e}^{-\frac{1}{2}\left(\frac{\epsilon_k^2}{k\sigma^2}\right)}}{\sqrt{2\pi k\sigma^2}},\\
       &= \prod_{k=1}^L \int \text{d}\epsilon_k \frac{1}{4}\left(2 + 2\cos\epsilon_k\right)\cdot  \frac{\text{e}^{-\frac{1}{2}\left(\frac{\epsilon_k^2}{k\sigma^2}\right)}}{\sqrt{2\pi k\sigma^2}},\\
       &= \frac{1}{2^L}\prod_{k=1}^L \left(1+\text{e}^{-\frac{k\sigma^2}{2}}\right).
    \end{split}
\end{equation}
The steps of this derivation is relatively insensitive to the parent distribution of $\epsilon$, and could be modified further if it were expected to be significantly different. A large number of simulations of the Draper adder with imprecise rotation gates were configured. Figure \ref{errorgraphs}a compares Equation (\ref{exact-error}) to these results. In Appendix \ref{app:error-ensemble} we also look at these errors in the context of components in Shor's algorithm. Figure \ref{errorgraphs}b illustrates the results of MPS-based simulations of an isolated modular adder compared with the predictions made by Equation (\ref{mod-error})
Equations (\ref{mod-error}) and (\ref{exact-error}) can be extrapolated to predict the performance of each respective arithmetic component in the regime of $L=2048$. Figure \ref{errorgraphs}c shows that for large $L$, the rotation error angle would need to be restricted to $\lesssim 5\times10^{-4} \:\text{rad}$ in order to deliver a result with appropriate fidelity.

%% file: adder_redesign.tex
\section{Resource-optimal redesign of the phase adder}
\label{sec:adder-redesign}

The truncation analysis so far has characterised errors, but made no attempt to address them. 
We aim to show that their systematic emergence can be targeted with corrections. 
The classical pre-computing in Fourier arithmetic means that not all parameters are as unknown as we have treated them. Correction gates based on the frequency of $1$s can be applied to eliminate a great deal of the known error source.
Suppose a known $a$ is added to an unknown $x$. 
From Section \ref{sec:analytic-trunc}, the asymptotic probability of finding a carry bit is $1/2$. 
Conditional on a given $a_j = 1$, however, the only way to \emph{not} have a carry bit is if $x_j=0$ and the $(j-1)^\text{th}$ bit is also not a carry. 
This increases the probability of error to $1 - 1/2\cdot1/2 = 3/4$. 
Applying corrective rotations contingent on each $a_j=1$ will resultantly eliminate $3L/8$ errors and introduce $L/8$ new errors; reducing the total number from $L/2$ down to $L/4$ in total. 
If the corrective gates are absorbed into the additive rotations, this increase in probability is at the cost of zero extra logical resources. 
In principle, a comprehensive model could be developed based on the full conditional error PMFs from knowledge of our numbers -- that is, computing the probability of a carry conditionally not just from the selected bit, but also from the value of its neighbours. 
For now, we will make corrective rotations relatively naively.\par
It was previously problematic every time a bit position summed up to an $\ell-$level carry bit. 
This would induced a carry factor of $p_\mathcal{N}^{\ell^2}$. 
If, with each $a_{j,k} = 1$, we applied a controlled corrective rotation of $\pi i/2^{\mathcal{N}+1}$ at a distance of $\mathcal{N}+1$ away, then an even number of $1$s would cancel the error entirely, and an odd number would multiply the register by just $p_\mathcal{N}^{1/4}$. 
With the reverse situation applied for subtractions, then the support of the effective sum on a single qubit is reduced from $(-2n,2n+2)$ to $[-1,1]$. 
The over-corrective errors would be orders or magnitude smaller and occur considerably less often.\par
\begin{figure}[h!]
    \centering
      \includegraphics[width=\linewidth]{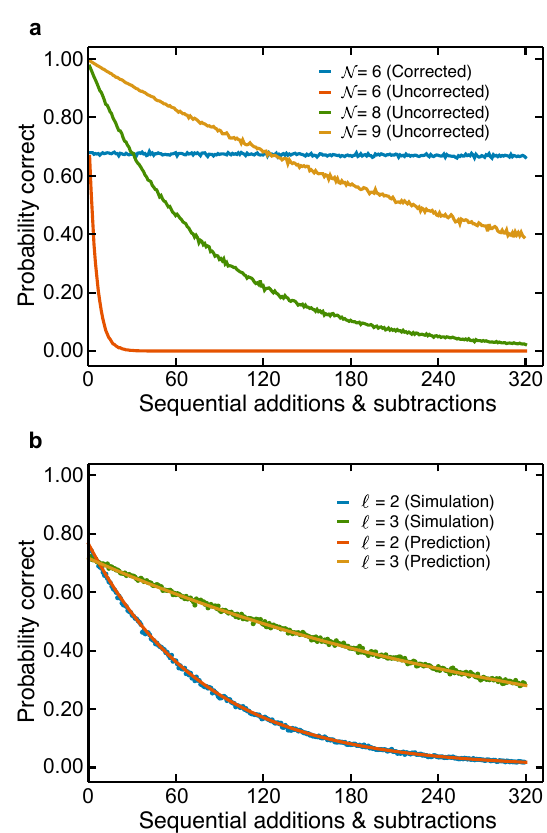}
\caption{\textbf{a} A comparison of the error-corrected adder shows that it starts with an initial probability very similar to an uncorrected $\mathcal{N}=6$, but sustains its fidelity for significantly longer in depth -- outperforming even truncated circuits three levels higher. $L=2000,$ $\mathcal{N}=6$ and $\ell=5$ with uncorrected versions at $\mathcal{N}=6,8,\:\text{and}\:9$ \textbf{b} Equation (\ref{corrected_prob}) is compared for $L=2048$, $\mathcal{N}=6$ to simulations for different $\ell$, showing good agreement.}
\label{correctederrors}
\end{figure}

The effective sum in the $j^\text{th}$ position is also influenced by the values of lesser significant bits. 
This means that the corrections so far will only account for the errors due to $c_i$; to correct for each lesser-significant bit $k$, we must apply rotations of $\pi i/2^{\mathcal{N}+k+1}$. 
Not all corrections are necessary. 
We will denote the parameter of number of corrections by $\ell$, wherein the next $\ell$ bits are corrected at a precision of up to $\pi i/2^{\mathcal{N}+\ell}$. 
Since all of the corrections can be collected into the single phase rotation, no extra logical resources are required.
Figure \ref{correctederrors}a shows a simulated comparison of a corrected truncated sequence of large-scale adders with uncorrected versions. 
This comes at a cost, where there is an initial error of $\mathscr{T}_{AS}^{\ell/6}$. The $\ell/6$ here follows from the proportion of rotation error introduced in the case of $(a_j,b_j) = (0,1)$ or $(1,0)$ (probability $1/2$), \emph{and} $x_j = 1$ or 0 respectively (probability $1/2$), \emph{and} no existing error chain (probability $2/3$).
The decay of the adder in depth is the same as an uncorrected truncated adder of level $\mathcal{N} + \ell$. 
For an adder truncated to level $\mathcal{N}$ and corrected to a level $\ell$, the probability of producing the correct result after $n$ additions and $n$ subtractions is:
\begin{equation}
    \label{corrected_prob}
    P_{\mathcal{N},\ell} = \left|\frac{1}{2}+\frac{1}{2}\text{e}^{\frac{i\pi}{2^\mathcal{N}}}\right|^{2\cdot \frac{L}{3}\frac{\ell}{6}}\cdot \left|\frac{1}{2}+\frac{1}{2}\text{e}^{\frac{i\pi}{2^{\mathcal{N}+\ell}}}\right|^{2\cdot \frac{n+1}{6}L}.
\end{equation}
Results of this prediction are compared with simulations in Figure \ref{correctederrors}b, to good agreement.
In Appendix~\ref{app:shor-trunc}, we explicitly run through the effects of stochastic error and truncation error in the context of Shor's algorithm (using the LNN circuit of Ref.~\cite{Fowler2004b}) and compute the success probabilites for various truncation regimes. These can be found summarised in Table~\ref{shor-trunc-comparisons} in the case of factoring RSA-2048, including a logical resource comparison to the uncorrected truncated adder. Remarkably, we find that with $\ell=11$ corrections on the adder rotation gates, the QFT can be truncated down to $\pi/64$ for $L=2048$, even though in the LNN implementation of Shor's algorithm there are $\mathcal{O}(16L^2)$ QFTs to be found. Alternatively, for uniform truncation (i.e. no corrections) and higher success probability, one can go only as coarse as $\mathcal{N}=17$.
\begin{table}
\centering

\begin{tabular}{@{}llcc@{}}
\toprule
$\mathcal{N}$ & $\ell$ & \thead{$\%$ Relative \\ Uncorrected Requirements} & Success Probability \\ \midrule
17 & 0 & 1 & 0.95 \\
6 & 11 & 0.35 & 0.32 \\
12 & 5 & 0.70 & 0.76 \\
13 & 4 & 0.76 & 0.89 \\ \bottomrule
\end{tabular}
\caption{A comparison of the performance and logical requirements of corrected vs. uncorrected truncated circuits of Shor's algorithm.}
\label{shor-trunc-comparisons}
\end{table}


%% file: cost_estimates.tex
\section{Surface code implementation cost analysis}
\label{sec:cost-analysis}
The error rates on Shor's algorithm are incredibly stringent, owing to its high depth.
It is likely therefore that an experimental implementation of the algorithm will only take place in the context of fault tolerant quantum error correction.
For this reason, we evaluate the resource cost of this redesigned circuit in the context of the surface code.\par 
Up to this point, savings have been quantified only in terms of unit-cost logical gates.
Here, we evaluate the QEC resources used by our truncated Fourier adder and make comparison to \textsc{Toffoli}-based circuits. In particular, we aim to estimate the number of distilled $T$-states (the logical $T$ gates consumed), as well as the number of raw magic states required in order to distill these $T$-states to an appropriately stringent error rate. This is for our truncated and corrected adder both on its own and in the context of Shor's algorithm, and for other state-of-the-art implementations. Although there are many cost considerations, a combination of logical qubit number and raw distillation costs provides a reasonable baseline comparison in a fault tolerant context.
The raw and distilled $T$-state estimates are based on the exact data provided by Campbell and O'Gorman in Ref.~\cite{rotation-magic-states} (supplemental material). Note that this does not take into account further optimisations or parallelisation which may be possible in individual circuit structure, but rather a basic conversion from logical gates into raw and distillable resources.\par
Each rotation $R_M$ by an angle $\pi/2^{M-1}$ is in the $M^\text{th}$ level of the Clifford hierarchy, and has a raw magic state cost of $\mathcal{C}(R_M, \eta)$ for desired logical error rate $\eta$. In particular, the higher levels of the Clifford hierarchy consume more resources to distill~\cite{Mooney2021costoptimalsingle}.
Assuming controlled rotation gates cannot be directly distilled -- $CR_z(\theta)$ decomposes into three single rotations of $\theta/2$, and two \textsc{CNot}s as: 
\begin{equation}
    R_z(\theta/2)_1\otimes R_z(\theta/2)_2\cdot\textsc{CNot}_{12}R_z(-\theta/2)_2\cdot \textsc{CNot}_{12}.
\end{equation}
This drives costs up one level, since controlled-$S$ gates require $T$ gates, controlled-$T$ gates require $\pi/8$ rotations, and so on. 
The rotations outside the \textsc{CNot} commute through the circuit: half of these are before the \textsc{Hadamard} in the QFT, and can be taken all the way to the left, and half of these are after and can be taken all the way to the right. The rotations at the right can be further commuted through the additive rotations, and will cancel out with the rotations similarly brought through the IQFT circuits. The rotations brought to the left cannot be omitted, however, in the context of many repeated QFT/IQFT combinations (such as in Shor's algorithm), these will cancel out in all instances except for the very first and very last (I)QFT. 

The third is trapped between both \textsc{CNot} gates, and cannot be commuted. Hence, the cost of the controlled phase rotations does not contribute extra distilled resources, but the raw resources will be higher because the rotations are twice as fine.
\begin{table*}[ht]
\centering
\resizebox{0.7\textwidth}{!}{
\begin{tabular}{@{}cccc@{}}
\toprule
 & Logical Qubits & Distilled Magic States & Raw Magic States \\ \midrule
Gidney~\cite{gidney-addition} & $4096$ & $8.19\times10^3$ & $4.13\times10^4$ \\
Häner \emph{et al.}~\cite{eth-adder} & $2049$ & $1.02\times 10^6$ & $1.33\times10^7$ \\
Draper~\cite{draper-adder} & $2048$ & $4.20\times 10^6$ & $5.31\times10^6$ \\
Truncated Draper $\mathcal{N}=7$ [this work] & $2048$ & $2.66\times10^4$ & $1.33\times10^6$ \\ \bottomrule
\end{tabular}}
\caption{Results comparing the resource requirements of different singular adders.}
\label{adder-costs}
\end{table*}
\begin{table*}[ht]
\centering
\resizebox{0.75\textwidth}{!}{
\begin{tabular}{@{}ccccc@{}}
\toprule
 & Logical Qubits & Distilled Magic States & Raw Magic States & Success Probability \\ \midrule
Gidney and Eker\aa~\cite{gidney2021factor} & $6189$ & $1.1\times 10^{10}$ & $3.98\times 10^{11}$ & 1 \\
Gidney~\cite{gidney-addition} & $6144$ & $8.00\times 10^{11}$ & $3.33\times 10^{13}$ & 1 \\
Häner \emph{et al.}~\cite{eth-adder} & $4098$ & $4.03\times10^{13}$ & $2.29\times10^{15}$ & 1 \\
Fowler \emph{et al.} (LNN)~\cite{Fowler2004b} & $4100$ & $1.40\times10^{14}$ & $4.01\times10^{15}$ & 1 \\
Truncated LNN: $17_0$ & $4100$ & $2.46\times10^{12}$ & $2.01\times10^{15}$ & 0.95 \\
Truncated LNN: $12_5$ & $4100$ & $1.78\times10^{12}$ & $1.16\times10^{15}$ & 0.76 \\
Truncated LNN: $6_{11}$ & $4100$ & $9.61\times10^{11}$ & $3.87\times10^{14}$ & 0.32 \\ \bottomrule
\end{tabular}}
\caption{Results comparing the resource requirements of different arithmetic regimes in the context of Shor's algorithm factoring RSA-2048. We can further optimise resources using coarser truncation at the cost of moderately reduced success probability.}
\label{Full-Shor-Cost-Analysis}
\end{table*}
Taking $H$ and $S$ gates to be free, then, for an exact $L$ qubit QFT/IQFT combination with an adder in the middle, the cost of the trapped target gates is $2\cdot \sum_{M=3}^{L-1}\mathcal{C}(R_{M},\eta)\cdot (L-M)$.
The cost of the commuted control qubit rotations is $2\cdot \sum_{M=3}^{L-1}\mathcal{C}(R_{M},\eta)$.
The cost of the commutable target gates is $L\cdot \mathcal{C}(R_{\langle M \rangle},\eta)$, where $\langle M \rangle$ is the average level required by an adder -- $\langle M \rangle = \mathcal{N}$, typically.
The respective total cost of an untruncated and truncated Draper adder, in terms of raw $T$ states, is therefore
\begin{equation}
    \label{draper-cost}
    \begin{split}
    &\mathcal{C}_\text{Draper} =\\ &\sum_{M=3}^{L+1}\left[2\cdot \mathcal{C}(R_{M},\eta)\cdot (L-M+2)\right]+L\cdot \mathcal{C}(R_{\langle M \rangle},\eta);
    \end{split}
\end{equation}
\begin{equation}
    \label{trunc-draper-cost}
    \begin{split}
        &\mathcal{C}_\text{Trunc} =\\ &\sum_{M=3}^{\mathcal{N}+2}\left[2\cdot \mathcal{C}(R_{M},\eta)\cdot (L-M+2) \right]+L\cdot\mathcal{C}(R_{\mathcal{N}},\eta).
    \end{split}
\end{equation}

\subsection{Comparison with Existing Toffoli Adders}
The distillation costs of rotations have recently favoured \textsc{Toffoli}-based circuits in the literature \cite{fourier-distill, cody-layered}. However, owing to the infancy of experimental quantum computing, the preferred resource focus is unclear. A reduction in qubit numbers may prove advantageous at the cost of more $T$-gates.
We aim to evaluate our construction in this context.\par

The adder construction in Ref.~\cite{eth-adder} is a \textsc{Toffoli}-based construction proposed to spatially compete with Fourier-based adders.
This required $\mathcal{O}(L\log_2 L)$ \textsc{Toffoli} gates, offering an increased gate count as cost for a decreased qubit count -- factoring in $2L+2$ qubits.
The most $T$-efficient known adder is a \textsc{Toffoli}-based adder which uses $4L+\mathcal{O}(1)$ $T$ gates on $2L$ qubits \cite{gidney-addition}.
In qubits and gates these are respectively the two most efficent \textsc{Toffoli} adders in the literature, and will be the two designs to which we compare our design.
The noise-level at which the comparison takes place will be set by the estimates made in Section \ref{sec:error-model}, unless there are significant discrepancies between the number of distilled $T$ gates -- in which case the noise threshold will be scaled by the ratio of resources.
For direct comparison's sake, we have taken all constructions to be a result of distillation of $T$ states. Note however, that direct distillation of \textsc{Toffoli} states is possibly more applicable to the \textsc{Toffoli}-based adders. A more fine-grained approach would be to compare the distillation schemes directly, but beyond the intent of this work.
In Refs.~\cite{eth-adder} and~\cite{gidney-addition}, we take, therefore $7$ and $4$ $T$ gates to produce a useful \textsc{Toffoli}, respectively.\par

A single Gidney adder requires $2L$ qubits and $4L+\mathcal{O}(1)$ distilled T states; a Häner adder requires $L$ qubits and $56L(\log_2 L - 2) + \mathcal{O}(1)$ distilled $T$ gates.
Under the pretext of $L=2048$, this is $8.2\times 10^3$ and $1.03\times10^6$ distilled $T$ gates.
A complete Draper adder requires $L+1$ qubits and $4.2\times 10^6$ distilled $T$ gates, meanwhile an $\mathcal{N}=7$ truncated Draper adder has $3.1\times 10^4$ distilled $T$ gates.
From our earlier analysis, we consider a distillation protocol for the truncated Draper and Gidney adders to be $\eta = 10^{-5}$.
Since the Häner and full Draper adders require two orders of magnitude more gates, we consider them in the regime of $\eta = 10^{-7}$. Note that distillation of angles in the full Draper adder as fine as the noise level is essentially free.
A complete comparison of raw resource requirements is summarised in Table \ref{adder-costs}.

A highly optimised Shor's algorithm circuit was recently published by Gidney and Eker\aa~\cite{gidney2021factor}, making use of windowed arithmetic to reduce the number of multiplications. We compare to this circuit as the state-of-the-art, as well as with standard modular exponentiation using the Gidney adder and the adder from H\"aner \emph{et al.} The latter is considered as well because many of the optimisations of Ref.~\cite{gidney2021factor} could also be applied to our truncated Draper adder, and so an unoptimised Gidney adder construction is more like-for-like with this work.\par 
In Ref.~\cite{gidney2021factor}, the quoted \textsc{Toffoli} figure to factor RSA-2048 is $2.7\times 10^9$. Assuming the same $T-$cost per Toffoli as in~\cite{gidney-addition}, this leaves the total number of distilled $T$ gates as $1.1\times 10^{10}$.
From the estimate in \cite{gidney-addition}, optimisations can be made to reduce the T count per Toffoli to $2.7$. 
This places the total number of distilled $T$ gates at $8.00\times 10^{11}$. 
The H\"aner \emph{et al}. construction in \cite{eth-adder} requires just $2L+2$ qubits, but a distilled $T$ count of $7\times 2L(32.01L^2(\log_2 L - 1) + 14.73L^2) = 4.03\times10^{13}$. 
Each modular adder in the LNN circuit construction consists of 2 QFT/IQFT combinations and 4 additions. 
There are $4L^2$ modular adders in the circuit, summing to a distilled magic state count of $1.40\times10^{14}$ for the full Draper adder; $9.61\times10^{11}$ for a $6_{11}$ truncated Draper circuit; $1.78\times10^{12}$ for a $12_5$ regime; and $2.46\times10^{12}$ for a $17_0$ circuit. 
From results in Section 6.1, we use a noise level for the Gidney and the truncated adder at $\eta = 10^{-12}$. 
Given that the full Draper and the H\"aner adders require a factor of $100$ more logical gates, we take these at a noise level of $\eta = 10^{-14}$. Finally, with Ref.~\cite{gidney2021factor} requiring the fewest logical resources, we distill using $\eta = 10^{-10}$. We then estimate the number of raw magic states required for distillation using the results of Ref.~\cite{magic-distillation-campbell}.

The results of this comparison are given in Table \ref{Full-Shor-Cost-Analysis}.
It is clear from Table \ref{Full-Shor-Cost-Analysis} that the truncated Fourier adder outperforms the resource requirements of~\cite{eth-adder}, and that the complete Draper circuit hosts the worst demands. $17_0$, $12_5$, and $6_{11}$ each require fewer raw magic states than the Häner construction, and respectively require a factor of $60$, $35$, and $12$ more raw magic states than the pure adder Gidney construction~\cite{gidney-addition}. Whether this is mitigated by the reduction in both logical and physical qubits will depend highly on how future experimental developments in quantum computing proceed. Note that using the results of Ref.~\cite{AQFT-nam}, the $T$ cost of our truncated arithmetic could be reduced further at the expense of more qubits. It appears possible that with further improvements to its error-correction scheme, coupled with more efficient distillation regimes, that the circuit with truncated arithmetic could outperform the \textsc{Toffoli} constructions in both $T$ count and number of qubits.

%% file: conclusion.tex
\section{Conclusion and Outlook}
\label{sec:conclusion}
In this article, we have provided a detailed study of the Draper QFT adder. 
The appeal of the Draper adder is in its low qubit requirements compared to other constructions and its elegant design, with the drawback of high gate count and circuit depth.
We first derived the exact effects of truncating out Fourier phase levels in the adder, showing that performing truncated addition with equal parts positive and negative rotations is the best way to minimise truncation errors and that truncations may be far coarser than previously expected. Adding to this, we investigated the effects of stochastic gate error in the circuit.
We then showed that, from knowledge of the error distribution, the QFT and IQFT -- whose gates compose the most substantial part of the Fourier adder -- can be made to be far more coarse than the additive rotations themselves. Given that the (I)QFT contributes the quadratic gate scaling in this implementation of arithmetic, the savings are significant.\par 
Using our modified construction, we looked at the cost of a qubit-efficient implementation of Shor's algorithm in a realistic surface code context, considering the factoring of RSA-2048 as an example. With no further attempt to optimise, we see that the raw resource requirements are comparable (or better than) those of \textsc{Toffoli} adder constructions -- pending the importance of total qubit number. Indeed, it is highly surprising that a 2048 qubit Shor's algorithm could survive with each QFT applied to a level of $\pi/64$. \par
More work could be conducted to design the circuit around mitigating these truncation errors or detecting their effects. For example, the corrective rotations could employ a more sophisticated conditional probability distribution. Or, in Shor's algorithm between modular additions the most significant qubit should always be set to zero, and between modular multiplications in Shor's algorithm the addition register should always be reset. This knowledge could be used to prevent truncation effects from propagating forward, permitting even coarser and more resource efficient implementations.
Given that arithmetic components play an integral role in classical computing, these results should find wide applicability in a variety of different quantum computing contexts beyond just Shor's algorithm.

\section*{Acknowledgments}
This work was supported by the University of Melbourne through the establishment of an IBM Quantum Network Hub at the University.
G.A.L.W. is supported by an Australian Government Research Training Program Scholarship. 
C.D.H. is supported through a Laby Foundation grant at The University of Melbourne.


%% file: appendices.tex
\section{Predicting the Average Truncation Effects for a Given L}
\label{sec:app-av}
Although we can deterministically compute the error incurred by a series of truncated adders when the numbers are known, this will not be the case in practice and it does not address the expected performance of the circuit component.
To this effect, we consider the sum of two unknown numbers and derive an expectation value for truncation's effects. 
Assuming that each qubit enters in a superposition of $(\ket{0}+\ket{1})/\sqrt{2}$ that bit's value can be treated as a Bernoulli random variable with equal probability.
This assumption is made, but does not confine us; our results allow for the respective probabilities of $0$s and $1$s to be changed if necessary.\par
We first wish to know the likelihood of encountering a string of $1$s of a given length. 
We will refer to these as \emph{success runs}. Let be $M_n^{(k)}$ the number of success runs with length $k$ or more in $n$ Bernoulli trials, the probability mass function (PMF) for this random variable is given \cite{bernoulli-prob} by
\begin{equation}
    P(M_n^{(k)}=x):= P(p,n,k,x) = \sum_{m=x}^{\lfloor\frac{n+1}{k+1}\rfloor} (-1)^{m-x}\binom{m}{x}p^{mk}q^{m-1}\left(\binom{n-mk}{m-1}+q\binom{n-mk}{m}\right).
\end{equation}
This machinery can be used to evaluate the probability loss incurred from the average addition of two numbers. 
Two quantities are required in order to calculate this: the average number of distinct carries within $L-\mathcal{N}-1$, and the average carry chain length. 
$P(M_n^1=x)$ communicates the probability of having $x$ runs in a given chain. 
From this, the average number of runs $R(p,n)$ in a given chain can be computed as:
\begin{equation}
    R(p,n) = \sum_{x=1}^{n}P(p,n,1,x)\cdot x.
\end{equation}
Equation (\ref{bernoulli-prob}) is a survival function -- that is, it generates the probability of having $x$ chains of length \textit{at least} $k$. 
To find the exact probability that we have $x$ chains of length $k$ we define
\begin{equation}
    P_2(M_n^{(k)}=x) := P_2(p,n,k,x) = P(p,n,k,x)-P(p,n,k+1,x).
\end{equation}
Further defining $S(p,n,k):=\sum_{x=0}^{n} P_2(p,n,k,x)\cdot x$, we then have the average number of length $k$ runs in $n$ trials. 
From this, the average length of a run can be computed:
\begin{equation}
    A(p,n) := \sum_{k=0}^{n}\frac{S(p,n,k)\cdot k}{R(p,n)}.
\end{equation}
A carry chain is propagated when either the sum $a_i+x_i$ is greater than or equal to one. 
The probability of this occurring is $3/4$, thus the average length of an error chain is given by $A(\tfrac{3}{4},L)$. 
All that remains to calculate is the average number of distinct carries $C(n)$ in the sum of two random numbers. 
The average number of distinct carries will be the total average number of initial carry events -- which is $L/4$, divided by the number of carries per error chain. 
This accounts for carry events hidden within an error chain without contributing to the loss of probability beyond the propagation of the chain.\par
By definition, a carry chain begins with a carry event. 
Conditional on the fact that that the chain propagates, each possibility for the rest of the chain is either $1 + 0, \: 0 + 1,\: \text{or}\: 1+1$, leaving a $\frac{1}{3}$ probability of carry bit. 
The total number of carry bits per error chain is therefore $1 + \frac{1}{3}\cdot\left(A(\tfrac{3}{4},n) - 1\right)$. 
This leaves us with the total number of distinct carry bits:
\begin{equation}
    C(L) = \frac{L}{4\cdot \left(1 + \frac{1}{3}\cdot\left(A(\tfrac{3}{4},L) - 1\right)\right)}.
\end{equation}
Finally, taking into account the fact that only carries within the lowest $L-\mathcal{N}-1$ bits will cause an error, we obtain the following expression for the expectation value of the correct probability given a truncation level $\mathcal{N}$ and a number size $L$:
\begin{equation}
    \mathscr{T}_A(L,\mathcal{N}) = \left|\frac{1}{2}\left(\exp\left(-\frac{i \pi}{2^\mathcal{N}}\right)+1\right)\right|^{2\times C(L-\mathcal{N}-1)\times A\left(\tfrac{3}{4},L-\mathcal{N}-1\right)}.
\end{equation}

\section{Computing the Average Truncation Effects for Arbitrary Sequences of Adders and Subtractors}
\label{sec:app-av-depth}
Here, we prove Equation \eqref{repeated-adder-probability}, an expression for the correct probability in a sequence of repeated adders and subtractors.
We begin by considering the exponent in Equation \eqref{as-correct-series}. The floor function can be alternatively written (for $x$ not an integer) as
\begin{equation}
    \lfloor x \rfloor = x - \frac{1}{2} + \frac{1}{\pi}\sum_{k=1}^\infty \frac{\sin(2\pi k x)}{k}.
\end{equation}
Then, the proportions in the exponent can be simplified as:
\begin{equation}
    \begin{split}
        \sum_{d=-2n}^{2n+2}(\lfloor d/2 \rfloor)^2\cdot \text{Pr}\left(D = d\right) &=\sum_{d = -2n}^{2n+2} \left(\frac{d}{2} - \frac{1}{2} + \frac{1}{\pi}\sum_{k=1}^\infty \frac{\sin(\pi k d)}{k}\right)^2\cdot \text{Pr}\left(D = d\right)\\
        &= \sum_{d = -2n}^{2n+2}\left(\frac{d^2}{4} - \frac{d}{2} + \frac{1}{4} + \frac{1}{\pi}\sum_{k=1}^\infty \frac{\sin(\pi k d)}{k}\left[d - 1 + \frac{1}{\pi}\sum_{k'=1}^\infty \frac{\sin(\pi k' d)}{k}\right]\right)\cdot \text{Pr}\left(D = d\right)
    \end{split}
\end{equation}
We then treat each member of the brackets in kind.
\begin{equation}
\begin{split}
    \sum_{d=-2n}^{2n+2} \left(\frac{d^2}{4}\right)\cdot \text{Pr}\left(D = d\right) &= \frac{1}{4}(\text{Var}[D] + \langle D \rangle^2)\\
    &= \frac{1}{4}\left(\frac{2n+1}{3} + 1\right)\\
    &= \frac{n+2}{6},
    \end{split}
\end{equation}
using the variance derived in Equation \eqref{variances} and the fact that $\langle D \rangle = 1$. Continuing:
\begin{equation}
    \sum_{d=-2n}^{2n+2} -\frac{d}{2}\text{Pr}\left(D = d\right) = -\frac{1}{2}\langle D \rangle = -\frac{1}{2},
\end{equation}
and
\begin{equation}
    \sum_{d=-2n}^{2n+2} \frac{1}{4}\text{Pr}\left(D = d\right) = \frac{1}{4}.
\end{equation}
Now 
\begin{equation}
\begin{split}
    \left\langle \sin(\pi k D)\right\rangle = \left\langle D\cdot \sin(\pi k D)\right\rangle = 0\quad \forall \:k\:\in\mathbb{N}
\end{split}
\end{equation}
for Gaussians with mean 1 (for a different mean, this value is nevertheless proportional to $\text{e}^{-\frac{\pi^2 k^2\sigma^2}{2}}$ and effectively zero), implying that 
\begin{equation}
    \sum_{d=-2n}^{2n+2}\frac{\sin(\pi k d)}{k}\left[d - 1\right]\text{Pr}\left(D = d\right) = 0.
\end{equation}
Finally:
\begin{equation}
    \begin{split}
         \frac{1}{\pi^2}\sum_{d = -2n}^{2n+2}\sum_{k=1}^\infty\sum_{k'=1}^\infty \frac{\sin(\pi kd)\sin(\pi k'd)}{kk'}\text{Pr}\left(D = d\right)&= \frac{1}{\pi^2}\sum_{d = -2n}^{2n+2}\sum_{k=1}^\infty\sum_{k'=1}^\infty \frac{\sin(\pi kd)\sin(\pi k'd)}{kk'}\delta_{kk'}\text{Pr}\left(D = d\right)\\
         &= \frac{1}{\pi^2}\sum_{d=-2n}^{2n+2}\sum_{k=1}^\infty \frac{\sin^2(\pi k d))}{k^2}\text{Pr}\left(D = d\right)\\
         &= \frac{1}{\pi^2}\sum_{d=-2n}^{2n+2}\sum_{k=1}^\infty \frac{\langle1/2 - \cos(2\pi k D))\rangle}{k^2}\\
         &= \frac{1}{2\pi^2} \sum_{k=1}^\infty \frac{1}{k^2}\\
         &= \frac{1}{2\pi^2} \frac{\pi^2}{6} = \frac{1}{12}.
    \end{split}
\end{equation}

Putting this all together, we have

\begin{equation}
    \begin{split}
        \sum_{d=-2n}^{2n+2}(\lfloor d/2 \rfloor)^2\cdot \text{Pr}\left(D = d\right) &= \frac{n+2}{6} - \frac{1}{2} + \frac{1}{4} + \frac{1}{12} \\ 
        &= \frac{n+1}{6}
    \end{split}
\end{equation}

\section{Linking the error model to magic state conventions}
\label{app:error-ensemble}
Here, we take our continuous error model -- rotation fluctuations around the $Z-$axis and put it in the form of a discrete ensemble-based error model. This form is the one most commonly used for QEC calculations, expressing an average probability of stochastic $Z-$ errors. Taking $\rho$ to be a Hermitian matrix allows us to express it as $\rho = \frac{1}{2} \left(\mathbb{I} + \textbf{n}\cdot\sigma\right)$. With our model definition we have our transformed density matrix: 
\begin{equation}
    \rho ' = \int P\left(\theta \right)R_z(\theta)\rho R_z^\dagger (\theta)\mathop{d\theta}
     = \frac{1}{2} \int P(\theta) R_z(\theta)\left[\mathbb{I} + n_x X + n_y Y + n_z Z\right]R_z^\dagger (\theta)\mathop{d\theta},
\end{equation}
where $P(\theta)$ is the PDF for the distribution of our rotation angles. Substituting and working through generates
\begin{equation}
    \begin{split}
        \rho ' &= \frac{1}{2}\left(\mathbb{I} + n_z Z\right) + \frac{1}{2} \int P(\theta) \mathop{d\theta}\left[n_x \begin{pmatrix} 0 & \text{e}^{-i\theta}\\ \text{e}^{i\theta} & 0\end{pmatrix} + n_y\begin{pmatrix} 0 & -i\text{e}^{-i\theta}\\ i\text{e}^{i\theta} & 0\end{pmatrix}\right].\\
    \end{split}
\end{equation}
Focusing on the second term:
\begin{equation}
    \begin{split}
        & \frac{1}{2}\int P(\theta)\mathop{d\theta} \left[n_x \cos\theta \begin{pmatrix} 0 & 1\\ 1 & 0\end{pmatrix} + n_x \sin\theta \begin{pmatrix} 0 & -i\\ i & 0\end{pmatrix} + n_y\cos\theta \begin{pmatrix} 0 & -i\\ i & 0\end{pmatrix}+n_y\sin\theta \begin{pmatrix} 0 & 1\\ 1 & 0\end{pmatrix}\right];\\
        &= \frac{1}{2} \int P(\theta)\mathop{d\theta} \left[X\left(n_x\cos\theta - n_y\sin\theta \right) + Y\left(n_x\sin\theta + n_y\cos\theta\right)\right].
    \end{split}
\end{equation}
For an even PDF, $\langle \sin\theta\rangle=0$, so our transformed density matrix is $$\rho' = \frac{1}{2}\left(\mathbb{I} + n_z Z + n_x\langle\cos\theta\rangle X + n_y \langle\cos\theta\rangle Y\right)$$
In the case of $P(\theta) = N(0,\sigma)$, the characteristic function is $\text{E}\left[\text{e}^{itx}\right]= e^{it\mu - \frac{\sigma^2t^2}{2}} = \text{e}^{- \frac{\sigma^2t^2}{2}}$. In order to find $\langle\cos\theta\rangle$ we can take $ \langle\cos\theta\rangle = \Re \text{E}\left[\text{e}^{ix}\right] = \text{e}^{- \frac{\sigma^2}{2}}$. So then $$\rho' = \frac{1}{2}\left(\mathbb{I} + n_z Z + n_x\text{e}^{-\frac{\sigma^2}{2}} X + n_y e^{-\frac{\sigma^2}{2}} Y\right).$$
This can be separated into
\begin{equation}
    \begin{split}
        \rho' &= \frac{1}{2}\left(\mathbb{I} + n_z Z + n_x\left(\frac{1}{2}e^{-\frac{\sigma^2}{2}} + \frac{1}{2}\right) X + n_y \left(\frac{1}{2}e^{-\frac{\sigma^2}{2}} + \frac{1}{2}\right) Y - n_x\left(-\frac{1}{2}e^{-\frac{\sigma^2}{2}} + \frac{1}{2}\right)X - n_y \left(-\frac{1}{2}e^{-\frac{\sigma^2}{2}} + \frac{1}{2}\right)Y\right),\\
        &= \frac{1}{2}\left(\mathbb{I} + n_z Z + n_x\left(\frac{1}{2}e^{-\frac{\sigma^2}{2}} + \frac{1}{2}\right) X + n_y \left(\frac{1}{2}e^{-\frac{\sigma^2}{2}} + \frac{1}{2}\right) Y + n_x\left(-\frac{1}{2}e^{-\frac{\sigma^2}{2}} + \frac{1}{2}\right)ZXZ + n_y \left(-\frac{1}{2}e^{-\frac{\sigma^2}{2}} + \frac{1}{2}\right)ZYZ\right),\\
        &= p\rho + (1-p)Z\rho Z,
    \end{split}
\end{equation}
where $p = \frac{1}{2}\left(\text{e}^{-\frac{\sigma^2}{2}}+1\right).$\\

%% file: LNN_Shor.tex
\section{Performance of the linear nearest-neighbour Shor's algorithm under truncation and error}
\label{sec:LNN-shor}
In the main text we developed a model to characterise a Fourier-based quantum adder with a limited set of rotation gates. 
Here, we explicitly evaluate and demonstrate its abilities in the context of Shor's algorithm. We use the circuit construction of Ref. \cite{Fowler2004b}. This exposes some unexpected caveats that require consideration.
We develop a model to characterise the implications a truncated adder has on the operation and resources of a complete run-through of Shor's algorithm on a large-scale quantum computer. We then include rotation errors to assess their effects on each of the components in the circuit.
\subsection{Analysis of the Circuit}
In order to evaluate the truncated Fourier adder's performance in Shor's algorithm, we first appraise the demands of modular exponentiation. 
The operation of exponentiation requires $\mathcal{O}(L^2)$ repeated additions. 
Equation (\ref{repeated-adder-probability}) from the encapsulates this aspect of the adder's performance both in size and depth. 
However, in order to make the operation \emph{modulo}, frequent \textsc{CNot} gates must be applied between the most significant bit of the ancilla register, and the MS qubit. 
This entangling operation engenders a loss of coherence within the calculation, preventing some errors from cancelling. \par
The full modular exponentiation circuit of Shor's algorithm involves $2L$ controlled modular multiplications, each of which has $L$ modular additions taking place on each of the working registers. 
Given that only half of each of these will be controlled, predictions of the circuit reduce to $L\cdot L/2 = L^2/2$ sequential modular additions. 
To estimate the overall circuit behaviour under truncation, we focus on extending Equation \eqref{repeated-adder-probability} to the repetition of modular additions.
It was shown in \cite{Fowler2004} that the period-finding QFT in Shor's algorithm could be truncated down to $\pi/64$. 
Since this is an accepted optimal result in the literature, and since this QFT sub-routine is common to all circuit implementations of Shor's algorithm, we will not discuss truncation of the QFT with respect to period-finding, only in the context of Fourier arithmetic. 
This will constitute the entire tool-set required in order to predict the effects of Fourier truncation on Shor's algorithm.
\subsection{Predicting a Single Modular Adder}
\begin{figure}[h]
    \centering
    \includegraphics[width=\linewidth]{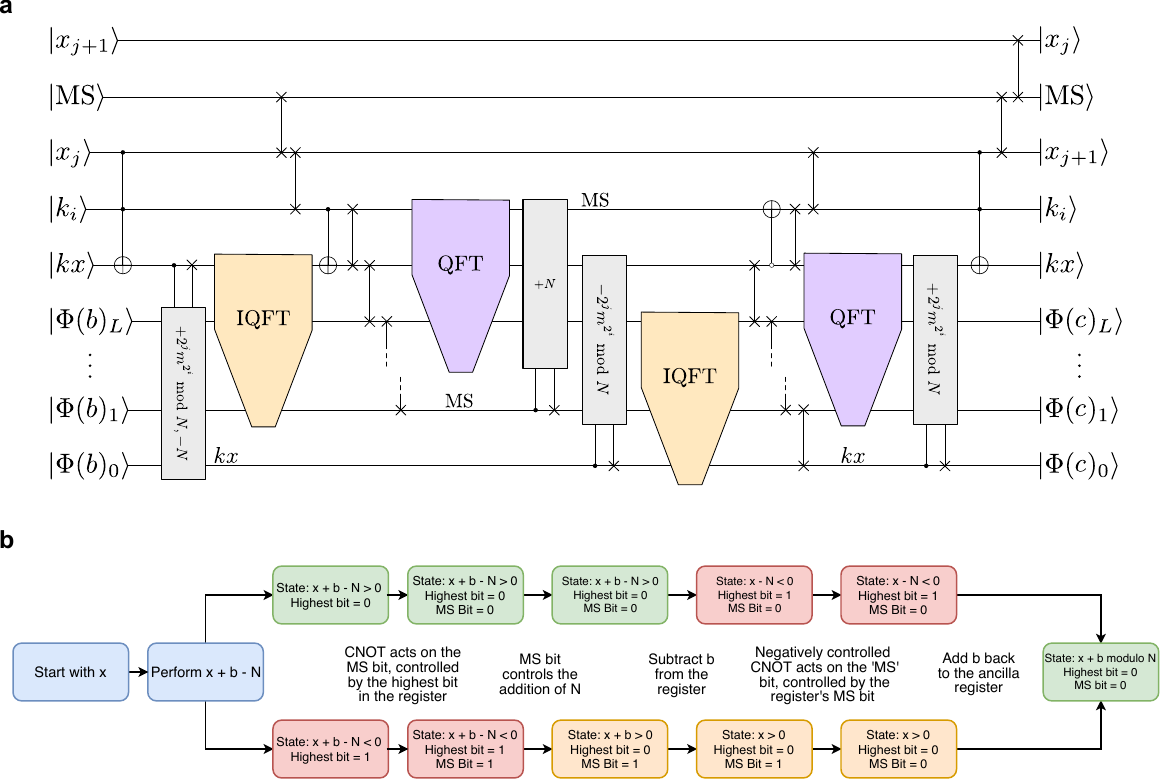}
    \caption{\textbf{a} Circuit diagram to compute a controlled modular addition as in the context of Shor's algorithm, redrawn from Ref~\cite{Fowler2004b}. \textbf{b} A flow chart depicting the logical components of the modular adder. The two pathways represent the controlled operations that take place when the register is in a positive or negative state, respectively. The colour green is for when the ancilla register is in the correct state $x+b\:\text{mod}\:N$, and $\ket{\text{MS}}=\ket{0}$; orange is for when the ancilla register is in the correct state, but the MS qubit is not reset; red is for the stages at which the ancilla register is in the incorrect state.}
    \label{fig:modadder_flow}
\end{figure}

The minimal component to Shor's algorithm is the modular adder. 
The characterisation of this single component will form the basis of evaluating the remainder of the circuit. 
The circuit diagram of the modular adder is given in Figure~\ref{fig:modadder_flow}a, with the logical flow in Figure \ref{fig:modadder_flow}b. Note here that the Toffolis are to be decomposed into any standard LNN set of gates, and the QFT circuits are as in Figure~\ref{draper}.
Summarily, it consists of a subtractor; a \textsc{CNot} between the MS qubit and the most significant qubit of the register; addition half of the time\footnote{If we consider our initial register $x$ to be a random number uniformly chosen in the interval $[0,N-1]$ and the number to which it is added, $y$, also chosen uniformly in the interval $[0,N-1]$, then the probability of $x+y < N$ is $1 - \frac{(N-2)^2}{2(N-1)^2}$. 
The average probability of subtracting off $N$: $\sum_{N=2^{L-1}}^{2^L}\left[1 - \frac{(N-2)^2}{2(N-1)^2}\right]\frac{1}{2^L-2^{L-1}}$ quickly approaches $\frac{1}{2}$}; a subtraction; another \textsc{CNot}; and an addition of the same number.\par
It is most convenient to separate the modular adder into two halves: the first half with the subtractor, \textsc{CNot} and adder; and the second half with an addition, \textsc{CNot}, and subtraction of the same number. 
The reason for this is that in the second half, the errors produced from truncation should cancel out entirely with the subtraction from the same number. 
Any probability leakage in this component of the modular adder can therefore be isolated out as the sole effect of the \textsc{CNot}. 
This is in contrast to the first half of the modular adder, wherein probability can also be lost due to pure truncation effects, and the two must be disentangled.
\subsection{Logical Decoherence with the CNot}
Entanglement has the potential to reduce phase coherence by introducing alternative states with which a computational pathway can be correlated. 
The effect of this can be characterised with a simple example. 
Consider a 6 qubit, $\mathcal{N}=4$ pathway of $14 + 13$, a \textsc{CNot}, followed by a subtraction of $13$. 
The initial addition, followed by a \textsc{CNot} produces the state
\begin{equation}
    \begin{split}
    |\psi\text{MS}\rangle &= \left(1-\text{e}^{-\frac{15i \pi}{16}}\right)|\text{MS}_1110011\rangle + \left(1+\text{e}^{-\frac{15i \pi}{16}}\right)|\text{MS}_0010011\rangle.\\
    \end{split}
    \end{equation}
Attempting to subtract $13$, the regular rotation procedure yields the state
\begin{equation}
\begin{split}
       \psi\rangle &= \left(1-\text{e}^{-\frac{15i \pi}{16}}\right)|\text{MS}_0\rangle\otimes \left(\left(1+\text{e}^{\frac{31i \pi}{16}}\right)|0\rangle + \left(-1+\text{e}^{\frac{31i \pi}{16}}\right)|1\rangle\right) \\
    &+\left(1+\text{e}^{-\frac{15i \pi}{16}}\right)|\text{MS}_1\rangle\otimes \left(\left(1+\text{e}^{\frac{15i \pi}{16}}\right)|0\rangle \left(-1+\text{e}^{\frac{15i \pi}{16}}\right)|1\rangle\right).
\end{split}
\end{equation}
Suppose there were no extra entanglement. 
Then the probability of obtaining the correct result $\ket{0}$ on the most significant bit is $$\frac{1}{4}\lvert\left(1-\text{e}^{-\frac{15i \pi}{16}}\right)\left(1+\text{e}^{\frac{31i \pi}{16}}\right) + \left(1+\text{e}^{-\frac{15i \pi}{16}}\right)\left(1+\text{e}^{\frac{15i \pi}{16}}\right)\rvert^2 = 1.$$ That is, we see that in the act of subtracting and adding the same number, we have no net error. However, when we consider the MS Bit, the phases can no longer constructively interfere and we are left with the probability of correct result $$\frac{1}{4}\lvert\left(1-\text{e}^{-\frac{15i \pi}{16}}\right)\left(1+\text{e}^{\frac{31i \pi}{16}}\right)\rvert^2 + \frac{1}{4}\lvert\left(1+\text{e}^{-\frac{15i \pi}{16}}\right)\left(1+\text{e}^{\frac{15i \pi}{16}}\right)\rvert^2 \approx 0.98097.$$
That is, the \textsc{CNot} causes us to have a final probability of $\left|1-\epsilon\right|^2 + \epsilon^2 \neq 1$ rather than $\left|1-\epsilon + \epsilon\right|^2 = 1$, preventing the phase errors from destructively interfering with each other. \par
\begin{figure}
\centering
\includegraphics[width=0.85\linewidth]{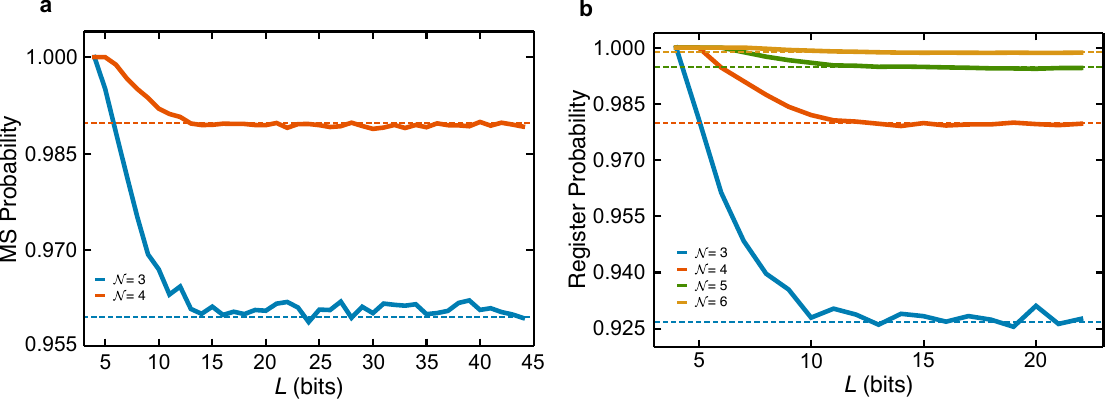}
\caption{Graphs to demonstrate the convergence of the interfering MS qubit. \textbf{a} Compares the MS correct-state probability with $p_\mathcal{N}$ after a single addition for increasing $L$. \textbf{b} Shows the register probability after an addition, \textsc{CNot}, and then subtraction. For all different truncation levels, the register converges to the value given in Equation (\ref{cnot-prob}).}
\label{converge_AS_graphs}
\end{figure}

This example illustrates how errors that \emph{should} recombine instead separate into an effectively mixed state. 
After a single addition, the number of states present is exponential in the number of errors. 
If an error occurs on a given bit in a particular pathway, that pathway will split into two further pathways. 
If both erroneous paths have the same state for the most significant bit ($\text{MS}=0$ or $\text{MS}=1$), then upon the subtraction of the same number, the errors will cancel out. 
These errors need not be considered. 
If, however, a given error separates into a superposition of $\ket{\text{MS}} = \ket{0}$ and $\ket{\text{MS}} = \ket{1}$, then after the application of the \textsc{CNot}, and an IQFT, the wave function will be in the form (neglecting normalisation):
\begin{equation}
    \ket{\psi} = \ket{\text{MS}_0}\otimes\left[\sum_j\left(1+\text{e}^{i\theta_j}\right)^{k_j}\ket{x_j}\right] + \ket{\text{MS}_1}\otimes\left[\sum_j\left(1-\text{e}^{i\theta_j}\right)^{c_j}\ket{y_j}\right].
\end{equation}
If we isolate out the target state, this is equal to:
\begin{equation}
\begin{split}
    \ket{\psi} &= \ket{\text{MS}_0}\otimes\left[\left(1+\text{e}^{i\theta_c}\right)^{k_c}\ket{\text{Correct}} + \sum_j\left(1+\text{e}^{i\theta_j}\right)^{k_j}\ket{\text{Incorrect}_j}\right]\\
    &+ \ket{\text{MS}_1}\otimes\left[\left(1-\text{e}^{i\theta_c}\right)^{k_c}\ket{\text{Correct}} + \sum_j\left(1-\text{e}^{i\theta_j}\right)^{k_j}\ket{\text{Incorrect}_j}\right].
\end{split}
\end{equation}
Following a negative phase rotation by the exact same amount and then subsequent IQFT, the amplitude of each state will be multiplied by its complex conjugate to give:
\begin{equation}
    \begin{split}
        \ket{\psi} &= \left(1+\text{e}^{i\theta_c}\right)^{k_c}\ket{\text{MS}_0}\otimes\left[\left(1+\text{e}^{-i\theta_c}\right)^{k_c}\ket{\text{Correct}} + \sum_j\left(1+\text{e}^{i\theta_j}\right)^{k_j}\ket{\text{Incorrect}_j}\right]\\
        &+ \left(1-\text{e}^{i\theta_c}\right)^{k_c}\ket{\text{MS}_1}\otimes\left[\left(1-\text{e}^{-i\theta_c}\right)^{k_c}\ket{\text{Correct}} + \sum_j\left(-1+\text{e}^{-i\theta'_j}\right)^{k_j}\ket{\text{Incorrect}_j}\right]
    \end{split}
\end{equation}
The final probability of obtaining the correct result, instead of being $1$, is now $$\left|(1+\text{e}^{i\theta_c})(1+\text{e}^{-i\theta_c})\right|^2 +\left|(1-\text{e}^{i\theta_c})(1-\text{e}^{-i\theta_c})\right|^2.$$ This is equal to $\left|(1+\text{e}^{i\theta_c})\right|^4 +\left|(1-\text{e}^{i\theta_c})\right|^4, $ which is the same as 
\begin{equation}
\label{cnot-prob}
    \text{Pr(MS Correct)}^2 + \left(1-\text{Pr(MS Correct)}\right)^2.
\end{equation}  
The errors -- which previously cancelled out -- are now in a mixed state corresponding to the different states of the MS qubit. 
In order to characterise the behaviour of the modular adder using our previous tools, the behaviour of the MS qubit evolution must be well-understood. 
It was shown in Section \ref{sec:trunc-depth} that the influence of chains of qubits is exponentially suppressed with each bit further down in the string. 
For this reason, and since only a single qubit is used for bookkeeping MS, the error approaches a constant rather than asymptotically growing with $L$.
As $L$ increases, the likelihood of any error depositing on the MS qubit approaches certainty. 
The convergent value of this MS state is therefore $p_\mathcal{N}$, with smaller order contributions from each subsequent qubit. 
Figure \ref{converge_AS_graphs}a illustrates this convergent behaviour by tracing out the density matrix of the MS qubit after a single adder for a range of values of $L$. 
This implies from Equation (\ref{cnot-prob}) that the effect of the \textsc{CNot} converges to multiplying the register out by $\mathcal{C}_S:=p_\mathcal{N}+(1-p_\mathcal{N})^2$. 
This conclusion is demonstrated in Figure \ref{converge_AS_graphs}b where the register probability is taken for a single addition, \textsc{CNot}, and then subtraction of the same number.

\subsubsection{Mixed State Fidelities with Addition and Subtraction of Different Numbers}
\begin{figure}
\centering
\includegraphics[width=0.85\linewidth]{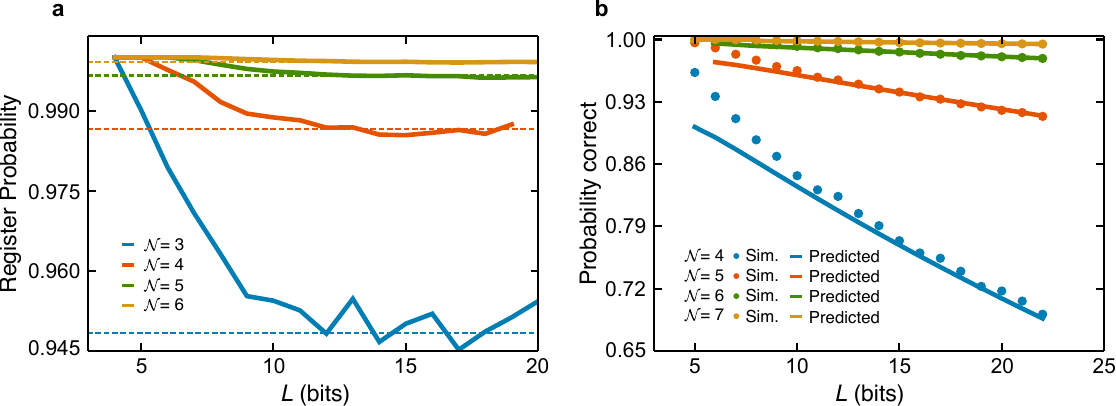}
\caption{\textbf{a} Shows the convergence the register probability of the addition and subtraction of different numbers with an interfering \textsc{CNot} with $\mathcal{C}_d$. This is as a ratio with the ideal case. \textbf{b} Compares the results of a single truncated modular adder with the prediction made in Equation (\ref{MA-equation}).}
\label{MA_predictions}
\end{figure}

The second component to a modular adder comprises the addition, \textsc{CNot}, and subtraction of a \emph{different} number.
Adder truncation multiplies the register by a factor $\propto 1-\theta^2$; mixed state errors introduce a factor $\propto 1-(\theta^2+\theta^2)$. 
In this case, a cancelled error performs \emph{worse} than a single carry error, and uncancelled errors are unaffected by the \textsc{CNot}. 
It was shown in Section~\ref{sec:analytic-trunc} that following an addition with a subtraction reduces $2/3$ of the errors. 
For this reason the effect of the \textsc{CNot} in the case of adding and subtracting different numbers will converge to $\mathcal{C}_D:=(2\cdot (p_\mathcal{N}+(1-p_\mathcal{N})^2) + 1)/3$. 
This behaviour is illustrated in Figure \ref{MA_predictions}a.
\subsubsection{Conclusion for a Single Modular Adder}
All of the components of the single modular adder are now completely characterised and can be pieced together. 
When the steps of the modular adder are subtraction $\rightarrow$ \textsc{CNot} $\rightarrow$ subtraction $\rightarrow$ \textsc{CNot}$ \rightarrow$ addition, the first \textsc{CNot} has no effect. 
Consequently, the total probability reduces to $\mathscr{T}_A\cdot \mathcal{C}_S$. 
If the steps are subtraction $\rightarrow$ \textsc{CNot} $\rightarrow$ addition $\rightarrow$ subtraction $\rightarrow$ \textsc{CNot} $\rightarrow$ addition, the probability can be expressed as $\mathscr{T}_{AS}\cdot \mathcal{C}_D\cdot \mathcal{C}_S$. 
The convergence of $\mathcal{C}_d$ and $\mathcal{C}_s$ means that for large $L$, the modular adder actually outperforms the adder. 
Given that each outcome is equally likely, the total average probability for a singular modular adder is therefore:
\begin{equation}
    \label{MA-equation}
   \mathscr{T}_{MA}(L,\mathcal{N}) = \frac{\mathcal{C}_S}{2}\left(\mathscr{T}_A(L) +\mathscr{T}_{AS}(L)\cdot \mathcal{C}_D\right).
\end{equation}

\subsection{Sequential Modular Adders}
\begin{figure}
\centering
\includegraphics[width=0.85\linewidth]{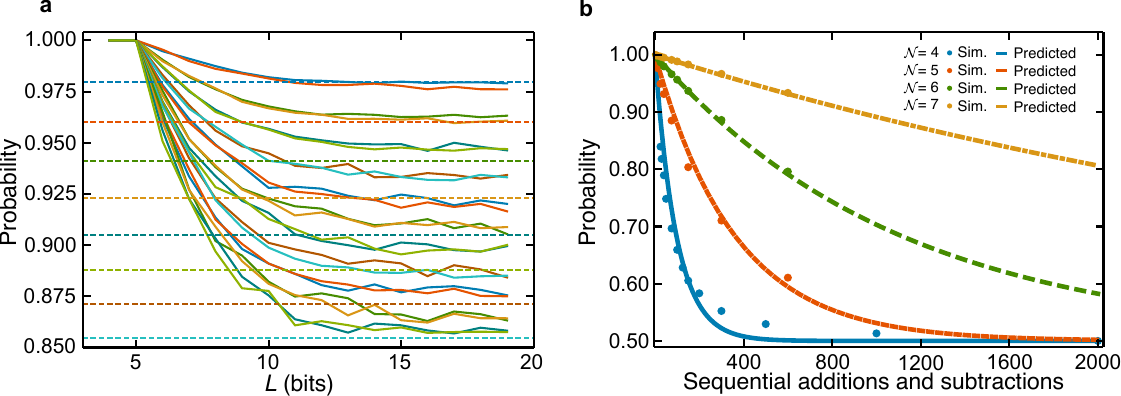}
\caption{\textbf{a} Compares the convergence of $1-22$ sequential adder/subtractor pairs with an interfering \textsc{CNot}. A drop in probability with every second addition is clearly seen. The dashed lines indicate the values that these probabilities would take if they multiplied out. \textbf{b} Compares equation \ref{cnot-depth-fidelity} with the simulations of a large number of sequential interfered additions and subtractions. For $\mathcal{N}=4$ \ref{cnot-depth-fidelity} over-estimates the decay slightly; any predictions made will err on the side of more error.}
\label{sequential_cnots}
\end{figure}

The scaling entanglement errors grows exponentially complex with $n$. 
Instead of a purely analytic expression, we provide an ansatz for the behaviour and then compare it with simulation results before making further predictions. 
Consider two applications of an addition and subtraction of a \textsc{CNot} in between. 
At this point, new entanglement errors could arise. 
Equally, however, the previous errors have the opportunity to disentangle from the MS qubit. 
For this reason, we expect every second application of this sequence to deliver a similar fidelity to the previous. 
Furthermore, multiple applications ought to retain the property of convergence with $L$. 
Finally, the register cannot be multiplied out with each application. 
The reason for this is that the interference of the MS qubit reduces the fidelity by producing a mixed state. 
The upper bound on its effect is when the state is \emph{maximally} mixed -- when $\left|\langle 1|\text{MS}\rangle\right|^2 = \left|\langle 0|\text{MS}\rangle\right|^2 = 0.5$, implying that the lower bound of fidelity due to MS decoherence is 0.5. 
The simulations in Figure \ref{sequential_cnots}a possess each of these traits.\par
\begin{figure}
\centering
\includegraphics[width=0.85\linewidth]{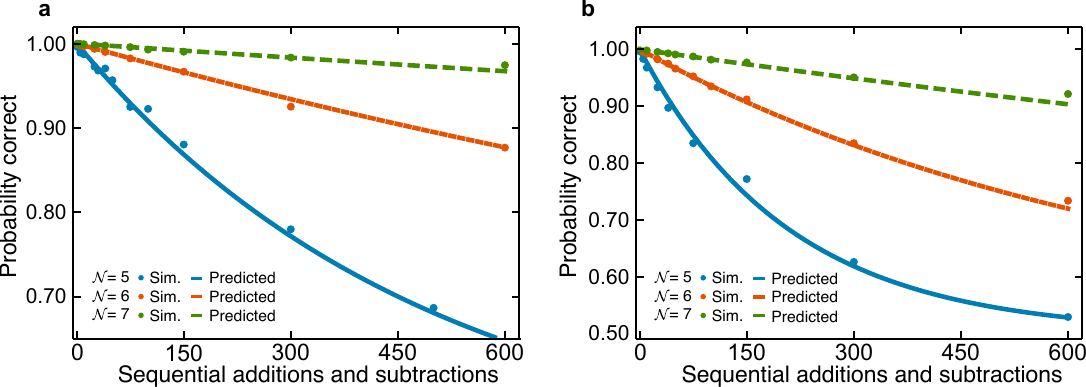}
\caption{\textbf{a} A comparison of Equation (\ref{asd-cnot-depth-fidelity}) with a large number of sequential interfered additions and subtractions of different numbers. These curves show good agreement with the data. Conducted for $L=18.$\textbf{b} A comparison of Equation (\ref{asds-cnot-depth-fidelity}) with a large number of $L=18$ sequential effective modular adders.}
\label{sequential_cnots-asd}
\end{figure}

We develop our model around the three properties deduced. 
Firstly, the limiting fidelity in depth must be convergent on 0.5. 
Secondly, there can be no $L$ dependence in the prediction. 
Finally, the fidelity must reduce on average with every second sequence. 
Based on this, our model for the fidelity $\mathscr{F}_s$ with depth of a sequence of $n$ additions and subtractions of the same number with interfering \textsc{CNot}s present is given by:
\begin{equation}
    \label{cnot-depth-fidelity}
    \mathscr{F}_s(n) = \frac{1}{2} + \frac{1}{2}\cdot \mathcal{C}_s^{\frac{n}{2}}.
\end{equation}
This prediction, Equation (\ref{cnot-depth-fidelity}) is compared with simulation results in Figure \ref{sequential_cnots}b. 
It appears to characterise the decay of the register's fidelity with depth.
Given that the decoherence plays the same role on the addition and subtraction of different numbers, it follows that the fidelity of this component, $\mathscr{F}_d$ is given by:
\begin{equation}
    \label{asd-cnot-depth-fidelity}
    \mathscr{F}_d(n) = \frac{1}{2} + \frac{1}{2}\cdot \mathcal{C}_d^{\frac{n}{2}}.
\end{equation}
Note that this characterises the effect of the \textsc{CNot} only; truncation decay is still given by Equation (\ref{repeated-adder-probability}). 
Figure \ref{sequential_cnots-asd} compares Equation (\ref{asd-cnot-depth-fidelity}) with the MPS simulations. 
The model as an exponential decay with $n/2$ appears to be correct. 
Finally, these two components can be combined to provide a prediction of a sequence of modular adders in the case of where the \textsc{CNot} interference has converged.
\begin{equation}
    \label{asds-cnot-depth-fidelity}
    \mathscr{F}_{ds}(n) = \frac{1}{2} + \frac{1}{2}\cdot \left(\mathcal{C}_d\cdot \mathcal{C}_s\right)^{\frac{n}{2}}.
\end{equation}
The results of this prediction can be seen in Figure \ref{sequential_cnots-asd}.

\section{Characterising the Fidelity of Shor's Algorithm}
\label{app:shor-trunc}
The complete modular exponentiation circuit in Shor's algorithm can largely be described by a sequence of repeated modular adders. 
A single register undergoes, on average, $L$ modular multipliers containing $L/2$ modular adders. 
Table \ref{shor-ma-comparison} compares the fidelity of Shor's algorithm with a sequence of $L^2/2$ modular adders. 
The figures agree very well. 
From this, we assert that the problem of determining the performance of Shor's algorithm under a truncated adder is exactly the problem of characterising the behaviour of sequential modular adders.
\begin{table}[]
\centering
\begin{tabular}{llll}
\toprule
$L$ (bits) & $\mathcal{N}$ & Simulated & Predicted \\ \midrule
5 & 3 & 0.630915 &  0.63462\\
6 & 3 & 0.442453  & 0.439068 \\
6 & 4 & 0.833247 & 0.824863 \\
7 & 3 & 0.237589 & 0.225458 \\
7 & 4 & 0.604293 & 0.599676 \\ \bottomrule
\end{tabular}
\caption{A comparison of truncated Shor's algorithm circuits with $L^2/2$ sequential modular adders.}
\label{shor-ma-comparison}
\end{table}
The final prediction for the performance of Shor's algorithm in the regime of a truncated Draper adder is consequently:
\begin{equation}
    \label{shor-predict}
    \begin{split}
            \mathscr{F}_\text{Shor} &= \left(\frac{1}{2} + \frac{1}{2}\cdot \left(\mathcal{C}_s\cdot\mathcal{C}_d\right)^{\frac{L^2}{4}}\right)\cdot p_\mathcal{N}^{L\frac{L^2+2}{24}},\\
            &=\left[\frac{1}{2}+\frac{1}{6}\left(\left(2(p_\mathcal{N}^2+(1-p_\mathcal{N})^2)+1\right)\left(p_\mathcal{N}^2+(1-p_\mathcal{N})^2\right)\right)^{\frac{L^2}{4}}\right]\cdot p_\mathcal{N}^{L\frac{L^2+2}{24}},
    \end{split}
\end{equation}
expressed in terms of the carry fidelity: $p_\mathcal{N} = \left|\frac{1}{2}+\frac{1}{2}\exp \left(-\frac{i \pi}{2^\mathcal{N}}\right)\right|^2$. 
Equation (\ref{shor-predict}) is plotted at $L=2048$ for different values of $\mathcal{N}$. 
We see that a sensible choice in order to retain an expectation of running the algorithm only twice is $\mathcal{N}\geq 15$.
\begin{figure}
\centering
\includegraphics[width=0.6\linewidth]{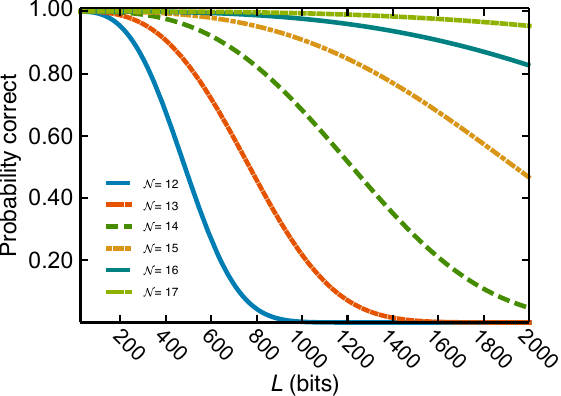}
\caption{A prediction of the effects of truncation on Shor's algorithm. For $L=2048$, it is clear that we must take $\mathcal{N}\geq14$ in order to keep the number of retries small.}
\label{shor-asymptotic}
\end{figure}


\subsection{Summary}
We have shown in our work that Shor's algorithm can be performed with a truncated Fourier adder at a level of just $\mathcal{N}=15$ and still be completed in polynomial time. This reduces the phase precision required from $\pi/2^{2048}$ down to $\pi/2^{15} \approx 1\times 10^{-4} $. Furthermore, it greatly reduces the required logical gate count for the whole circuit. For a truncation level $\mathcal{N}$, the gate count of a QFT is reduced by $\frac{(L-\mathcal{N}-1)(L-\mathcal{N})}{2}$. There are $16L^2 + 4L+1$ QFTs in a circuit. The new gate count is therefore:
$$L^3(46 + 16\mathcal{N}) + L^2 \left(-8\mathcal{N}^2 - 4\mathcal{N} + \frac{2325}{2}\right) + L\left(5 - \mathcal{N} - 2\mathcal{N}^2\right) - \frac{\mathcal{N}}{2} - \frac{\mathcal{N}^2}{2} - 2$$
Since we have reduced the resource count of the QFT from being quadratic in $L^2$ to linear, we are able to completely eliminate the $8L^4$ term in our circuit gate count. For $L=2048$ this amounts to a saving of $1.399\times10^{14}$ gates, leaving approximately $1.22\times 10^{12}$ -- less than $1\%$ of the required logical gates.


\section{Robustness of Other Arithmetic Circuit Components to Stochastic Phase Error}
\label{app:shor-error}
The Draper adder comprises larger arithmetic components in quantum circuits through a known number of repeated phase rotations. Consequently, the number of $\epsilon$ errors in Equation (\ref{error-expression}) is scaled linearly by the number of QFTs, IQFTs, and adders. For example, the most significant bit in equation (\ref{exact-error}) encounters the sum of precisely $(L-2)/2$ errors in the QFT, $1$ error in the adder, and $(L-2)/2$ in the IQFT, giving a total $\epsilon \stackrel{d}{=} N(0,(L-1)\cdot\sigma^2)$. This generalises to $j(L-2)$ errors from $j$ QFT and IQFT combinations, and $c$ errors from $c$ phase rotations. The distribution of $\epsilon$ through the application of this arbitrary number of components is therefore $j(L-2) + c$. Equation (\ref{exact-error}) can be modified to yield a performance expression for a Fourier-based arithmetic circuit containing $j$ QFTs and IQFTs, and $c$ adders. This is made explicit in the equation: 
\begin{equation}
\label{general-error}
    \langle P \rangle(j,c,L)=\frac{1}{2}\prod_{k=1}^L \left(1+\text{e}^{-\frac{(j\cdot (k-2)+c)\sigma^2}{2}}\right).
\end{equation}
This allows for a general prediction to be made of any arithmetic circuit making use of Fourier-based arithmetic. In particular, we focus on the components of Shor's algorithm.
\subsection{Errors in the Modular Adder}
A simple analysis of an isolated Mod Adder shows that there is $1$ QFT and $1$ IQFT for transitioning the ancilla register into the Fourier basis, as well as 2 QFTs and 2 IQFTs with which to perform arithmetic operations. Furthermore, there are 4 adders, one of which is truly controlled only half the time. This yields $j=3$ and $c=3.5$ for a single modular adder\footnote{Extrapolating to $n$ sequential modular adders would find $j=2n+1$ and $c=3.5n$, since the QFT and IQFT for the ancilla register need only to be applied once.}. As well as substituting in these values, it must be accounted for that an $L-$bit modular adder takes place on an $L+1$ qubit ancilla register. The expression for the probability is therefore:
\begin{equation}
\label{mod-error}
    \langle P \rangle_{MA}(L)=\frac{1}{2}\prod_{k=1}^{L+1} \left(1+\text{e}^{-\frac{(3\cdot (k-2)+3.5)\sigma^2}{2}}\right).
\end{equation}
Figure \ref{errorgraphs}b illustrates the results of MPS-based simulations of an isolated modular adder compared with the predictions made by Equation (\ref{mod-error})

Equations (\ref{mod-error}) and (\ref{exact-error}) can be extrapolated to predict the performance of each respective arithmetic component in the regime of $L=2048$. Figure \ref{errorgraphs}c shows that for large $L$, the rotation error angle would need to be restricted to $\lesssim 5\times10^{-4} \:\text{rad}$ in order to deliver a result with appropriate fidelity. 
\subsection{Performance and Resource Requirements of Shor's algorithm}
\begin{figure}[h!]
    \centering
    \includegraphics[width=\linewidth]{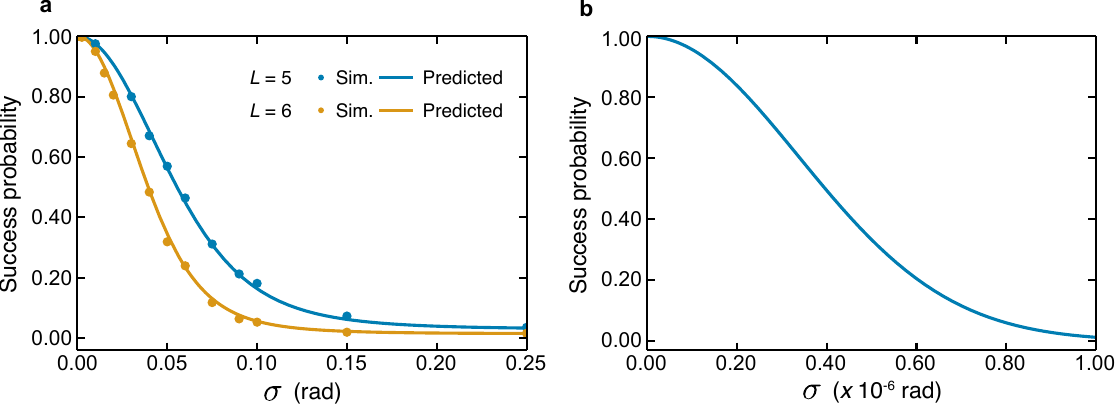}
    \caption{\textbf{a} Compares the predictions made by Equation (\ref{shor-error-eqn}) with the average result of 250 erroneous simulations of Shor's algorithm. \textbf{b} extrapolates these predictions to produce the expected fidelity of Shor's algorithm for $L=2048$}
    \label{fig:shor-errors}
\end{figure}

Equation (\ref{general-error}) extrapolates straightforwardly to the entirety of the modular exponentiator in Shor's algorithm. In each modular adder, the QFTs are carried out insensitive to the control of the addition. In comparison, on average only half of the additions are controlled and so only half yield errors. In a modular multiplier, each register is subject to $L$ modular adders, and a QFT. This implies that $j=2L+1/2$ and $c=3.5\cdot L/2$. However, each qubit will also control the inverse modular addition, inheriting each of those controlled errors as well. Therefore $c=3.5\cdot L$. If a modular multiplier is uncontrolled, then each QFT takes place on an empty ancilla register, introducing no errors. As a result, on average only $L$ modular multipliers operate. The final fidelity coefficients for the $L$ qubit register is $j=2L^2+L/2$ and $c=3.5\cdot L^2$. Consequently, the expression for the fidelity of a complete modular exponentiation circuit is given by:
\begin{equation}
\label{shor-error-eqn}
    \langle P \rangle_{ME} = \frac{1}{2}\prod_{k=1}^L \left(1+\text{e}^{-\frac{((2L^2+L/2)\cdot (k-2)+3.5L^2)\sigma^2}{2}}\right).
\end{equation}
A comparison of full-circuit simulations with Equation (\ref{shor-error-eqn}) are shown in Figure \ref{fig:shor-errors}a. We also extrapolate these results to provide an estimate of the circuit behaviour for $L=2048$. Examining Figure \ref{fig:shor-errors}b suggests that a reasonable target error rate in order to receive a correct result within two trials is $\sigma \approx 4\times10^{-7}$ rad. In terms of the noise on our rotation gates, this translates to $\eta = 4\times 10^{-14}$